\documentclass[%
 aip, superscriptaddress, amsmath,amssymb,
preprint,%
]{revtex4-1}

\usepackage{graphicx}% Include figure files
\graphicspath{{figures/}}
\usepackage{epstopdf}
\usepackage{dcolumn}% Align table columns on decimal point
\usepackage{bm}% bold math
\usepackage{xcolor}
\usepackage[utf8]{inputenc}
\usepackage[T1]{fontenc}
\usepackage{mathptmx}
\usepackage[version=3]{mhchem}

\begin{document}

\preprint{Applied Physics Letters \textbf{119}, 083904 (2021). DOI:10.1039/d1ee00650a}

\title{Counter-balancing light absorption and ionic transport losses in the electrolyte for integrated solar water splitting with III-V/Si dual-junctions}
 
\author{Moritz Kölbach}
 %\altaffiliation[Also at ]{Physics Department, XYZ University.}%Lines break automatically or can be forced %with \\
\affiliation{Universität Ulm, Institute of Theoretical Chemistry,  Lise-Meitner-Str.~16, 89069 Ulm, Germany}
\author{Ciler Özen}
\affiliation{ 
Helmholtz-Zentrum Berlin für Materialien und Energie GmbH, Institute for Solar Fuels, Hahn-Meitner-Platz 1, 14109 Berlin, Germany %\\This line break forced with \textbackslash\textbackslash
}
\author{Oliver Höhn}%
\author{David Lackner}%
\author{Markus Feifel}%
\affiliation{Fraunhofer Institute for Solar Energy Systems ISE, Heidenhofstraße 2, 79110 Freiburg, Germany}
\author{Fatwa F.~Abdi}%
\affiliation{ 
Helmholtz-Zentrum Berlin für Materialien und Energie GmbH, Institute for Solar Fuels, Hahn-Meitner-Platz 1, 14109 Berlin, Germany %\\This line break forced with \textbackslash\textbackslash
}
\author{Matthias M.~May}
\email{matthias.may@uni-ulm.de}
\affiliation{Universität Ulm, Institute of Theoretical Chemistry,  Lise-Meitner-Str.~16, 89069 Ulm, Germany}

%\date{\today}
%\linenumbers
\begin{abstract}

Recently, significant progress in the development of III-V/Si dual-junction solar cells has been achieved. This not only boosts the efficiency of Si-based photovoltaic solar cells, but also offers the possibility of highly efficient green hydrogen production via solar water splitting. Using such dual-junction cells in a highly integrated photoelectrochemical approach and aiming for upscaled devices with solar-to-hydrogen efficiencies beyond 20\%, however, the following frequently neglected contrary effects become relevant: (i) light absorption in the electrolyte layer in front of the top absorber and (ii) the impact of this layer on the ohmic and transport losses. Here, we initially model the influence of the electrolyte layer thickness on the maximum achievable solar-to-hydrogen efficiency of a device with an Si bottom cell and show how the top absorber bandgap has to be adapted to minimise efficiency losses. Then, the contrary effects of increasing ohmic and transport losses with decreasing electrolyte layer thickness are evaluated. This allows us to estimate an optimum electrolyte layer thickness range that counterbalances the effects of parasitic absorption and ohmic/transport losses. We show that fine-tuning of the top absorber bandgap and the water layer thickness can lead to an STH efficiency increase of up to 1\% absolute. Our results allow us to propose important design rules for high-efficiency photoelectrochemical devices based on multi-junction photoabsorbers.

\end{abstract}

\maketitle

Hydrogen produced from water and sunlight offers the potential to significantly contribute to the decarbonisation of the energy sector on a global scale.\cite{Krol_perspective_photoelectrochemical_energy_storage_2017, Ardo_pathways_electrochemical_solar_hydrogen_2018} 
One possible route towards solar hydrogen is photoelectrochemical (PEC) water splitting. In short, a dual-junction photoabsorber immersed in an electrolyte captures the incident sunlight, generates a photocurrent as well as a photovoltage, and drives the oxygen and hydrogen evolution reactions at the respective semiconductor-electrolyte interfaces. Despite decades of research, however, no material system was demonstrated that fulfils all of the following requirements for a commercially viable PEC system: (i) a lifetime on the time scale of years, (ii) high abundance of the absorber and catalyst materials, and (iii) a high solar-to-hydrogen (STH) efficiency. The latter is especially important, since efficiency becomes a key factor determining the hydrogen production cost, when the balance of system (BOS) and land costs shift away from the materials costs. Moreover, techno-economic analyses imply that only highly efficient PEC water splitting might compete with the technological more mature approach of powering an electrolyser by photovoltaics via the grid.\cite{Shaner_technoeconomic_analysis_renwable_H2_2016} The current record PEC device with respect to efficiency is based on a GaInP/GaInAs dual-junction cell, reaching 19\% STH.\cite{Cheng_tandem_2018} However, the relatively low stability and the high-cost of the required GaAs-substrate are currently preventing practical applications.\cite{Pinaud_hydrogen_production_facilities_2013, Tournet_review_III-V_water_splitting_2020}

One possible way to significantly reduce the cost of III-V-based devices is to replace the GaAs-substrate with Si that also acts as a bottom absorber.\cite{Cariou2018, Tournet_review_III-V_water_splitting_2020, Chen_assessment_GaPSb-Si_photoelectrochemical_cells_2021} Indeed, demonstrated efficiencies of III-V/Si multi-junction photovoltaic solar cells have significantly increased in the recent years. Also for integrated solar water splitting, there has been an increased interest for this approach over the last years.\cite{Vanka_InGaN-Si_water_splitting_2020, Chen_assessment_GaPSb-Si_photoelectrochemical_cells_2021} In 2018, Cariou et al. achieved a photovoltaic conversion efficiency of 33.3\% with a wafer-bonded two-terminal GaInP/GaAs//Si solar cell under AM1.5G illumination, which was further improved to 34.1\%.\cite{Cariou2018,Lackner_2020} Only recently, a new record of 35.9\% was achieved using a wafer-bonded two-terminal GaInP/GaInAsP//Si cell.\cite{ISE_press_release} Direct growth of the III-V cell(s) on top of a Si bottom cell offers potential cost and scalability benefits, but is also more challenging due to defects at the III-V/Si interface, which is why achieved efficiencies are still lower than those reported for the wafer-bonding approach.\cite{Feifel_directly_grown_III-V_Si_cell_2019, Feifel_2021} These developments pave the way for the development and fabrication of III-V/Si dual-junction cells for solar water splitting that promise similar high efficiencies as recent PEC record devices. Moreover, there has been significant progress in Perovskite/Si dual-junction devices\cite{Al-Ashouri1300, Koehnen_2021}, which are also considered potential candidates for highly efficient solar water splitting. It should, however, be noted that the longterm stability of direct or integrated approaches is still a major challenge that needs to be addressed before III-V/Si or Perovskite/Si photoelectrochemical dual-junction cells can reach large-scale commercial applications.\cite{High_efficiency_book_chapter_2018}

With STH efficiencies above 20\% within reach, a number of effects become relevant that are frequently neglected in lower-efficiency devices, but cannot be ignored when approaching the physical limits. In a dual-junction, two-terminal cell, the two absorbers are connected in series and the overall efficiency is determined by the absorber with the lowest current (current matching). Hence, the efficiency is highly sensitive to changes in the solar spectrum. In any PEC device, the incident light has to pass through a -- typically aqueous -- electrolyte before reaching the absorber. Since water absorbs near-infrared light, the effective illumination spectrum onto the cell deviates from the AM1.5G spectrum. Efficiency losses are therefore unavoidable and can be even more emphasised when the bandgaps of the multi-junction cells are perfectly matched to the AM1.5G spectrum instead of to the effective spectrum.\cite{Parkinson_efficiency_stability_PEC_devices_1984, Murphy_water_filter_water_splitting_2006, Doescher_sunlight_water_2014, High_efficiency_book_chapter_2018} An obvious strategy to minimise the parasitic absorption in the electrolyte is to decrease the water layer thickness in front of the absorber. However, with decreasing electrolyte thickness, ohmic and transport losses may in-turn decrease the efficiency representing a typical non-linear optimisation problem, leading to a global maximum of the theoretical efficiency as a function of the electrolyte layer thickness. Due to the current-matching condition in a monolithic dual-junction, this will then directly impact the ideal bandgaps of the photoabsorbers.

In this work, we deconvolute the effects of the electrolyte layer thickness on the efficiency of a III-V/Si dual-junction device. Therefore, we initially investigate the influence of the electrolyte layer thickness on the maximum achievable solar-to-hydrogen efficiency and show how the top absorber bandgap has to be adapted to minimise efficiency losses under idealised conditions. Next, we model the ohmic and transport losses with decreasing electrolyte layer thickness. We use experimental III-V//Si dual-junction device data (as the performance target for directly grown III-V/Si) and combine the effects of parasitic absorption and ohmic/transport losses. We hereby show that fine-tuning of both the top absorber bandgap and the water layer thickness can enable an absolute STH efficiency increase of up to 1\%.

\begin{figure*}[ht!]
\includegraphics{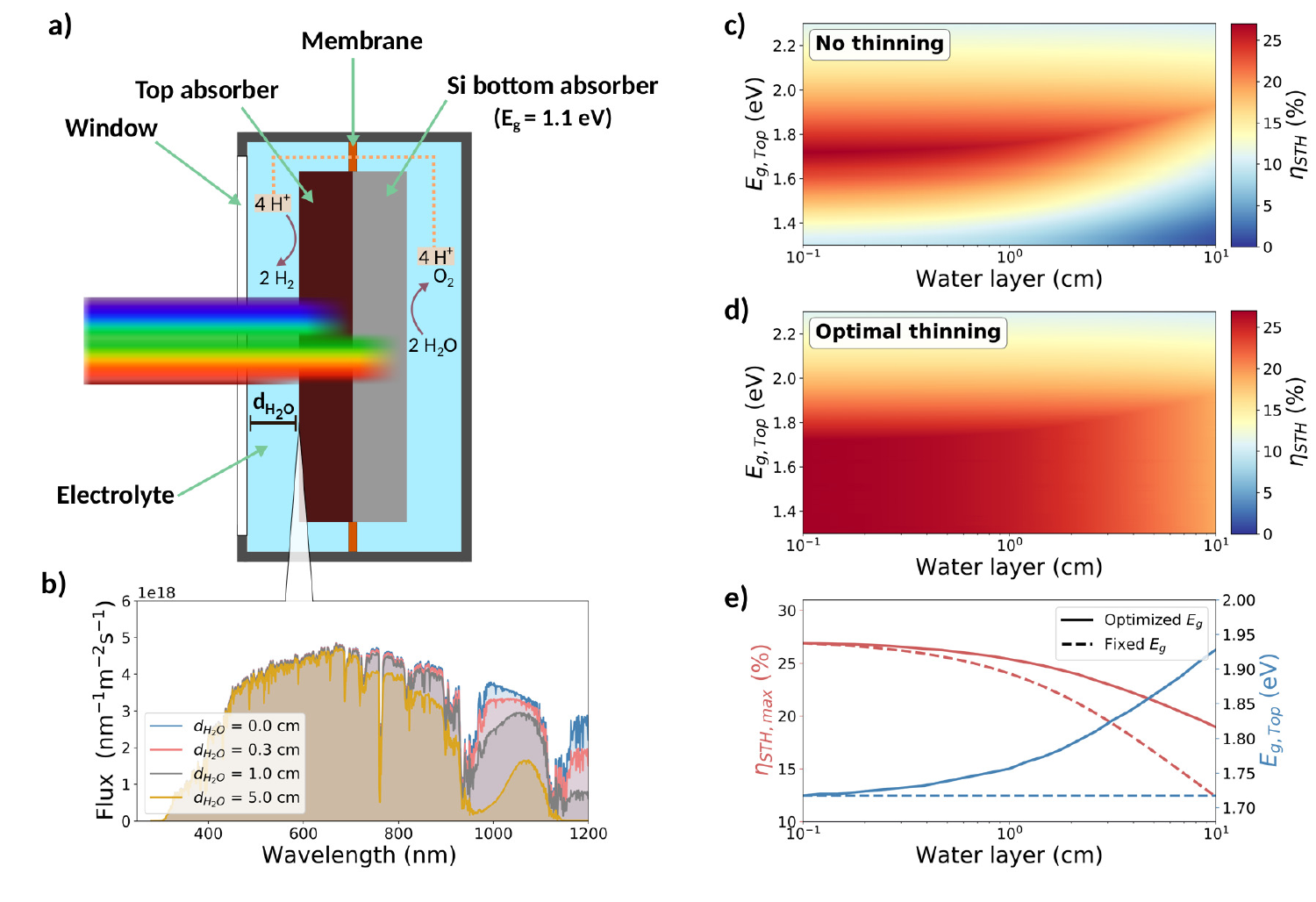}
\caption{\label{Figure_1} (a) Sketch of a dual-junction cell for PEC solar water splitting indicating the absorption of the incident light in the electrolyte. (b) Effective AM1.5G spectra (reference data from the American Society for Testing and Materials\cite{Gueymard_2004}) modified by the water layer (absorption data from Refs.\cite{Kou_1993, Pope_absorption_water_visible_1997}) with different thicknesses hitting the top absorber of the dual-junction cell. (c) STH efficiencies modelled for an ideal case as a function of the water layer thickness and the top absorber bandgap from the detailed balance limit and Pt- and \ce{IrO_x}-catalyst characteristics without thinning of the top absorber, and (d) with optimal thinning. (e) Maximum achievable STH efficiency for a fixed (dashed lines) and optimised (solid lines) top absorber bandgap as a function of the water layer thickness. The ohmic cell resistance is assumed to be constant in all calculations (i.e. water layer thickness-independent).
}
\end{figure*}

Fig.~\ref{Figure_1}(a) shows a sketch of a monolithic dual-junction PEC device with Silicon as a bottom absorber, indicating the near-infrared light absorption in the electrolyte under AM1.5G illumination. The effective spectra that reach the cell as a function of the water layer thickness in the relevant wavelength-range (E$_{g, Si}$ = 1.1\,eV, i.e.~1127\,nm) is shown in Fig.~\ref{Figure_1}(b). Even a thin water layer of 0.3\,cm decreases the effective intensity for wavelengths $>980$\,nm. When the water layer thickness is 5\,cm, the intensity is drastically decreased for wavelengths $>700$\,nm. The absorption coefficient stays finite also for lower wavelengths,\cite{Pope_absorption_water_visible_1997} but the effect becomes negligible as water layers of serveral cm are practically not reasonable, also due to the resulting weight of the device. However, it emphasises the need for a reliable benchmarking protocol for the characterisation of PEC multi-junction devices in the lab.\cite{May_benchmarking_PEC_2017} 

To initially only assess the influence of light absorption in the electrolyte layer on the device efficiency, a constant (i.e.~water layer thickness-independent) ohmic cell resistance is assumed in our calculations shown in Figs.~\ref{Figure_1}(c), \ref{Figure_1}(d), and \ref{Figure_1}(e). This ideal case scenario is modelled with our open-source Python package YaSoFo\cite{May_yasofo-data_2020} is based on the following conditions: highly efficient Pt- and \ce{IrO_x} catalysts (see Figure S1 for IV-characteristics), neglected concentration overpotentials, open circuit voltages obtained from the detailed balance limit, IV solar cell characteristics following the single diode equation,  and an operating temperature of 25\,$^\circ$C (see SI Table S1 for full list of input parameters and the YaSoFo documentation for full model description\cite{May_yasofo-data_2020}). Fig.~\ref{Figure_1}(c) shows the STH efficiency as a function of the top absorber bandgap and the water layer assuming that all photons with an energy higher than the bandgap are absorbed and contribute to the photocurrent. The same plot in which the thickness of the top absorber is allowed to be optimised from growth to be not fully absorbing (i.e.~ ``thinned''), so that more photons can reach the Si bottom absorber to ensure current matching conditions is shown in Fig.~\ref{Figure_1}(d). The maximum theoretical efficiency decreases with increasing water layer thickness from 26.9\% (0.1\,cm), 26.3\% (0.4\,cm), and down to 18.9\% (10\,cm). This represents an intrinsic efficiency loss in the photoelectrochemical solar water spitting approach. The red dashed line in Fig.~\ref{Figure_1}(e) shows the higher efficiency losses, when the top absorber bandgap is not adapted to the water layer thickness (blue dashed line). Interestingly, Figs.~\ref{Figure_1}(c) and \ref{Figure_1}(d) imply that there are two ways to minimise these losses and reach the maximum theoretical efficiency: (i) decrease the thickness of the top absorber so more photons can reach the Si bottom absorber, or (ii) increase the bandgap of top absorber (solid blue line in Fig.~\ref{Figure_1}(e). While both approaches have the same maximum achievable efficiency for a given catalyst performance, the latter would increase the photovoltage allowing for higher ohmic losses in the device, or the use of less catalyst loading or catalysts with a lower activity, respectively.

\begin{figure}[h!]
\includegraphics{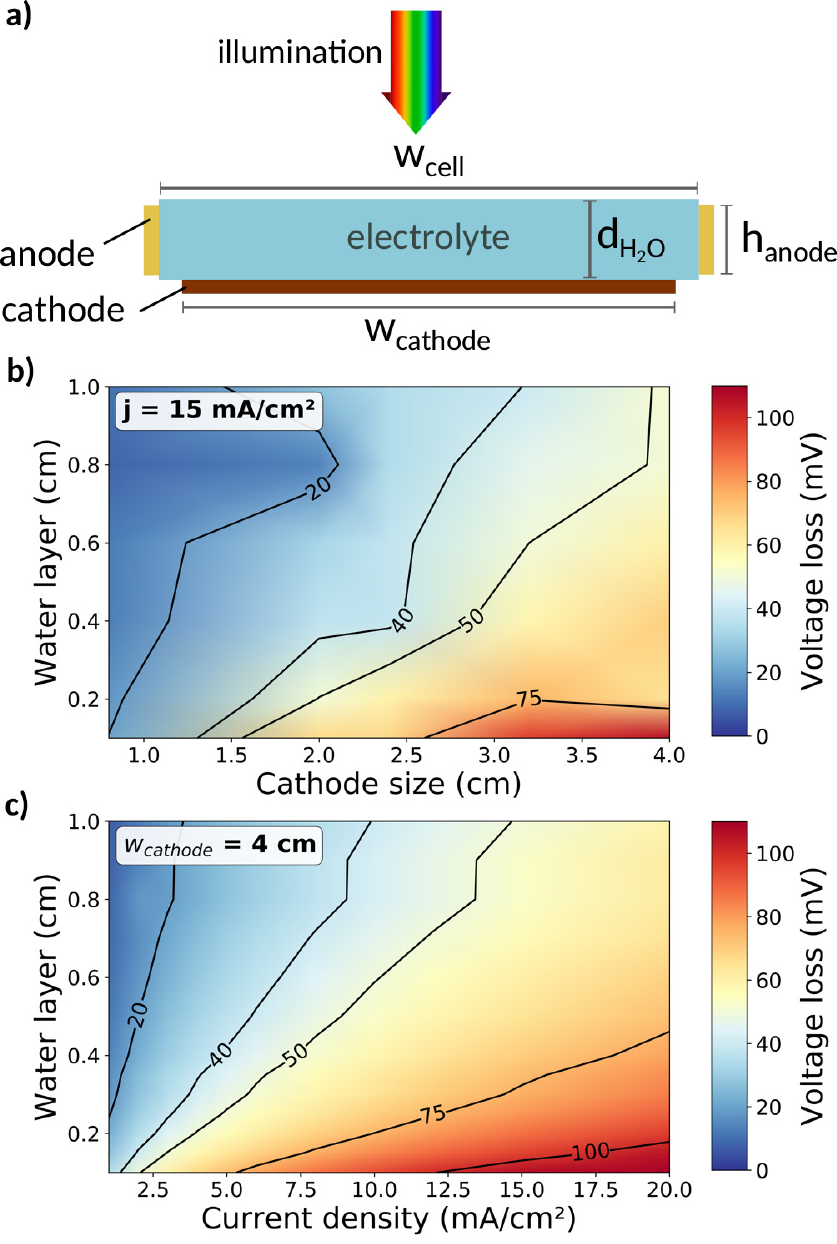}%
\caption{\label{fig:voltage_losses} (a) Sketch of the idealised 2-D cell geometry used for the modelling indicating the dimensions and arrangement of the cathode, the anodes, and the electrolyte. (b) Sum of the ohmic and concentration voltage losses for a constant current density of 15\,mA/cm$^2$ as a function of the cathode size simulated with COMSOL. (c) Respective calculations varying the current density for a constant cathode size of 4\,cm. The individual contributions of the ohmic and concentration losses are shown in the SI Figure S2.}
\end{figure}

To model the influence of the water layer thickness on the voltage losses, we used a simplified 2-D cell geometry as shown in Fig.~\ref{fig:voltage_losses}(a). The calculations assume a stagnant 1\,M \ce{HClO4} electrolyte, a more realistic operation temperature of 40\,$^\circ$C, no membranes, and anodes that are placed on the sides of the cell. Note that this highly idealised cell geometry does not represent a practical water splitting device (e.g. no safe product separation). However, it gives a first impression on the voltage losses associated with a thin water layer. Note that the size of the gas bubble plume\cite{Obata_bubbles_2021} creates another boundary condition for the minimum thickness of the water layer, but this is currently neglected in the model. To asses the ohmic losses and concentration overpotentials in the cell, the steady-state conservation, Nernst-Planck (diffusion and migration), as well as the concentration-dependent Butler-Volmer equation were solved employing the finite element method in COMSOL Multiphysics (see SI supplementary note 1 and Table S2 for more details and input parameters, respectively). 

As expected, our calculations show that the voltage losses (sum of the ohmic and concentration overpotentials) increase with decreasing electrolyte thickness and increasing cathode size (see Fig.~\ref{fig:voltage_losses}a). This is caused by the reduced cross-section of the conductive water layer leading to higher ohmic losses and mass-transport limitations. For example, the voltage loss doubles from 51\,mV to 106\,mV while decreasing the water layer thickness from 1\,cm to 0.1\,cm for a fixed current density of 15\,mA/cm² and a cathode size of 4\,cm. For smaller cathode sizes, the voltage losses also double, but the absolute losses are lower. These results emphasise the need for PEC device configurations that allow short ion path lengths. Since the absolute voltage losses increases with increasing cathode size, this is especially important for upscaled devices. The influence of the current density shown in Fig.~\ref{fig:voltage_losses}(c) also reveals the expected trend: The voltage losses increase with increasing current density and decreasing water layer thickness. For a cathode size of 0.4\,cm, the voltage loss increases from 58\,mV to 116\,mV with decreasing water layer thickness from 1\,cm to 0.1\,cm for a current density of 20\,mA/cm². Note that the individual contribution of ohmic and concentration losses are similar in the considered parameter space (see SI Figure S2).

\begin{figure*}
\includegraphics{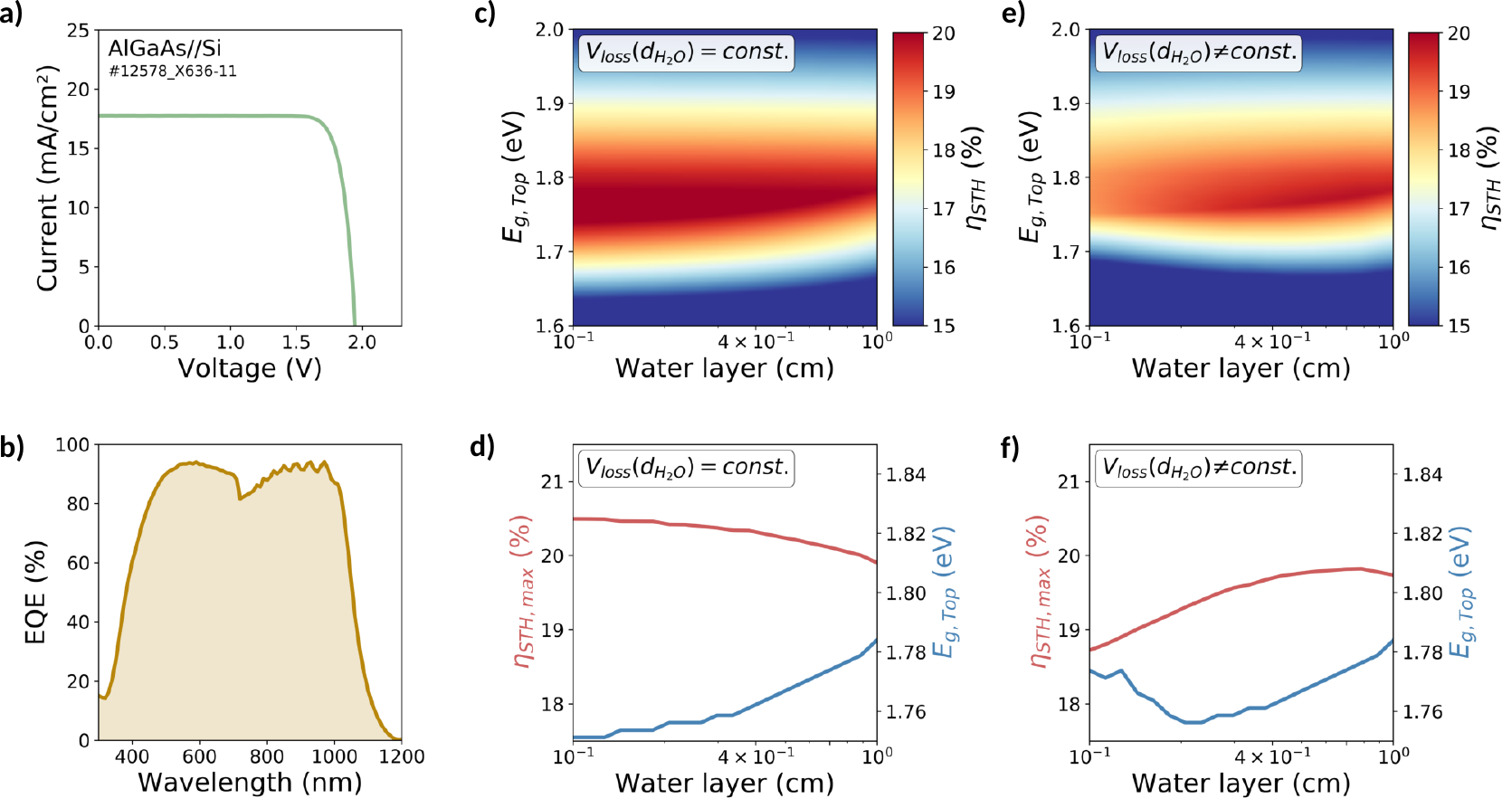}
\caption{\label{Figure_3} (a) IV characteristics under AM1.5G illumination of a AlGaAs//Si solar cell prepared by wafer bonding measured in the Fraunhofer ISE CalLab. (b) Sum of the EQEs of the two subcells measured at Fraunhofer ISE CalLab. (c) STH efficiencies modelled by YaSoFo\cite{May_yasofo-data_2020} as a function of the water layer thickness and the top absorber bandgap assuming a constant voltage loss. (d) Extracted maximum efficiency and the associated top absorber bandgap. (e), (f) Respective calculations considering the voltage losses modelled with COMSOL shown in Fig.~\ref{fig:voltage_losses}(c) for a cathode size of 4\,cm. All calculations were performed without the possibility of top absorber thinning. Full list of input parameters and the resulting 2-electrode water splitting IV characteristics are shown in the SI Table S1 and Figs. S3/S4, respectively.}
\end{figure*}

In order to determine the optimal condition, we now combine the two effects of the electrolyte layer: parasitic light absorption and the voltage losses caused by the ohmic and concentration overpotentials. For a more realistic analysis, experimental state-of-the-art III-V//Si dual-junction device data instead of the previously employed detailed balance limit is used. Figs.~\ref{Figure_3}(a) and~\ref{Figure_3}(b) show the IV characteristics under AM1.5G illumination and the external quantum efficiency (EQE) spectrum of an AlGaAs//Si solar cell, respectively. The AlGaAs top absorber ($E_{g}$ = 1.75\,eV) was joined with the Si bottom absorber via wafer bonding resulting in photovoltaic conversion efficiencies of up to 29.1\% (see Ref.~\cite{Lackner_2020} for experimental details). In the calculations shown in Fig.~\ref{Figure_3}(c-f), the experimental IV and EQE data is used as an input to inter alia account for parasitic absorption, recombination losses, as well as ohmic and finite shunt resistances in the absorber (see supplementary note 2 and Table S1 for full list of input parameters). The assumed operating temperature of 40\,$^\circ$C is implemented via the temperature coefficient of the open circuit voltage. Note that for an even more realistic device modelling, the optics of the total stack (i.e. air/window/water/catalyst/protection layer/absorber) as well as the exact cell geometry would have to be considered. However, this is out of the scope of the current study and is left for future work.

 Fig.~\ref{Figure_3}(c) shows the STH efficiency based on the experimental AlGaAs//Si cell characteristics as a function of the top absorber bandgap (no thinning) and the water layer thickness without taking the additional voltage losses due to a thinner water layer into account. The extracted maximum STH efficiency and the associated top absorber bandgap is illustrated in Fig.~\ref{Figure_3}(d). As expected, the calculations show a similar trend as those shown in Fig.~\ref{Figure_1}(c) based on the detailed balance limit. However, a lower maximum STH efficiency of 20.5\% is achieved for the smallest considered water layer of 0.1\,cm. Figs.~\ref{Figure_3}(e) and ~\ref{Figure_3}(f) show the respective calculations considering the additional voltage losses caused by the thinned water layer from Fig.~\ref{fig:voltage_losses}(c). The effect is clearly visible. The overall maximum achievable STH efficiency has shifted away from the lowest considered water layer thickness of 0.1\,cm to a value of around 0.7\,cm. In this thickness region, the effects of the parasitic absorption and voltage losses are counterbalanced. Figs.~\ref{Figure_3}(e) and~\ref{Figure_3}(f) reveal the important conclusion that fine-tuning of both the top absorber bandgap and the water layer thickness can lead to an absolute STH efficiency increase in the order of 1\%.

\begin{figure}[h!]
\includegraphics{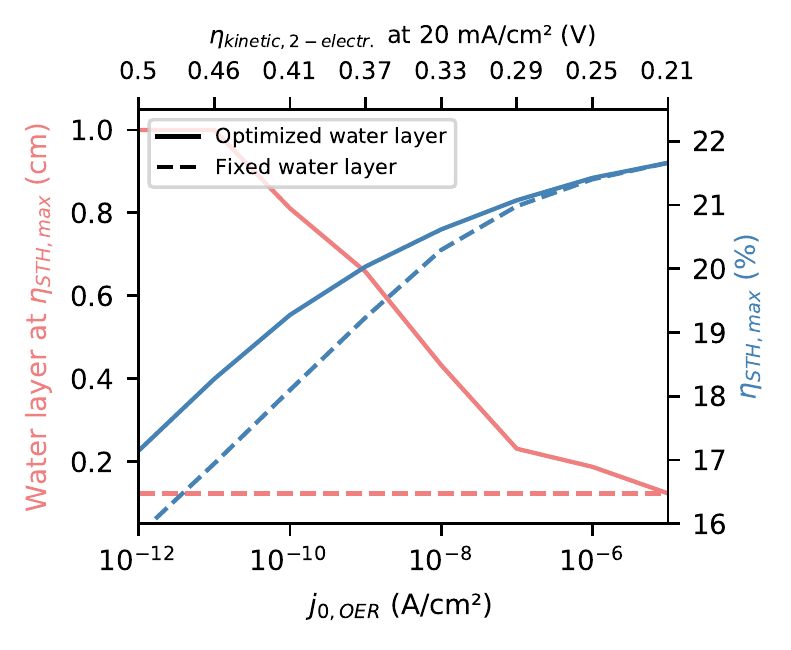}% 
\caption{\label{fig:max_water_cat} Water layer thickness (red solid line) for which the overall maximum STH efficiency (blue solid line, right y-axis) is achieved as a function of the OER exchange current density used as an figure of merit for the overall 2-electrode catalyst performance. The top x-axis indicates the kinetic overpotential to achieve a 2-electrode water splitting current density of 20\,mA/cm². The blue dashed line represents the respective maximum efficiency values for a fixed water layer of 0.1\,cm. }
\end{figure}

Another parameter that influences the trade-off between parasitic light absorption and ion transport losses in the water layer, which was not discussed until now, is the 2-electrode water splitting catalyst performance (see Fig.~\ref{fig:max_water_cat}). In the model, the catalyst performance was varied by changing the exchange current density of the OER catalyst (x-axis), while keeping the exchange current density of the HER catalyst constant. The resulting kinetic overpotential to achieve a 2-electrode water splitting current density of 20\,mA/cm$^2$ is indicated on the upper x-axis as a more tangible value. The red and blue solid lines show the optimised water layer thickness (left y-axis) and the corresponding overall maximum STH efficiency (right y-axis) as a function of the catalyst performance, respectively. For clarification, these values correspond to the maximum of the STH efficiency vs.~water layer thickness plot shown in Figure~\ref{Figure_3}(f). Note that each optimised water layer thickness also has a corresponding optimised top absorber bandgap (not shown). Furthermore, the distribution of the current density flowing in the electrolyte is assumed to be not affected by the exchange current density (i.e. the voltage losses shown in Fig.~\ref{fig:voltage_losses}(c) are independent of the catalyst performance).

When the overall efficiency is limited by current matching in the dual-junction cell and not by the catalysis, the solar cell generates more voltage than required (larger top absorber bandgap to allow more photons to reach the Si bottom absorber). In other words, the system operates at potentials below the MPP of the solar cell. This extra voltage, which is otherwise transformed into heat, can be used to counterbalance the voltage losses caused by a thin water layer (see Fig.~\ref{fig:voltage_losses}). Hence, maximum STH efficiencies are reached at low water layer thicknesses due to a lower parasitic absorption for a very good catalyst performance (right-hand side of Fig.~\ref{fig:max_water_cat}). If, on the other hand, the catalyst performance is limiting and defines the magnitude of the top absorber bandgap, every additional ion transport voltage losses will lower the maximum achievable STH efficiency. Hence, the maximum efficiency is reached at elevated water layer thicknesses. Note that for a realistic device, it is more likely that the catalyst performance will be the limiting factor. For comparison, the dashed blue line shows the respective maximum STH efficiency when the water layer is fixed, i.e.~not adapted to a reduced catalyst performance (dashed red line). The maximum achievable efficiency gains are again in the order of 1\% (absolute), when the water layer is optimised with respect to the catalyst performance.

In summary, we modelled and deconvoluted the effect of the water layer on the maximum achievable STH efficiency of a III-V/Si dual-junction device for PEC solar water splitting. We showed that fine-tuning of both the top absorber bandgap and the water layer thickness to counterbalance the effects of parasitic absorption and voltage losses can lead to an STH efficiency increase of up to 1\%. Moreover, our study emphasises the need to explore device designs that minimise the ohmic and transport losses associated with a thin water layer. This work lays the foundation for the development of a realistic PEC device model. Extending our calculations with experimentally obtained optical properties of the total stack (air/window/water/catalyst/protection layer/absorber), a practical and upscaled cell geometry, and including the influence of convection will be the subject of follow-up work. From a broader perspective, our results give important insights into the challenges of designing any highly efficient multi-junction PEC system, also beyond solar water splitting.\\

This work was supported by the German Bundesministerium für Bildung and Forschung (BMBF), project ``H2Demo'' (03SF0619A–K), the European Union’s Horizon 2020 research and innovation programme within the project SiTaSol (grant agreement no. 727497), as well as the German Research Foundation (DFG), project no.~434023472.

\section*{Data Availability}

The base code used for this study is available open source on the Zenodo repository under DOI:10.5281\textbackslash zenodo.1489157, specialised functions used here will be made available upon publication of the manuscript.  All further data, including the description of the model and all the input parameters, is available in the supplementary materials.

\providecommand{\noopsort}[1]{}\providecommand{\singleletter}[1]{#1}%


\begin{thebibliography}{25}%
\makeatletter
\providecommand \@ifxundefined [1]{%
 \@ifx{#1\undefined}
}%
\providecommand \@ifnum [1]{%
 \ifnum #1\expandafter \@firstoftwo
 \else \expandafter \@secondoftwo
 \fi
}%
\providecommand \@ifx [1]{%
 \ifx #1\expandafter \@firstoftwo
 \else \expandafter \@secondoftwo
 \fi
}%
\providecommand \natexlab [1]{#1}%
\providecommand \enquote  [1]{``#1''}%
\providecommand \bibnamefont  [1]{#1}%
\providecommand \bibfnamefont [1]{#1}%
\providecommand \citenamefont [1]{#1}%
\providecommand \href@noop [0]{\@secondoftwo}%
\providecommand \href [0]{\begingroup \@sanitize@url \@href}%
\providecommand \@href[1]{\@@startlink{#1}\@@href}%
\providecommand \@@href[1]{\endgroup#1\@@endlink}%
\providecommand \@sanitize@url [0]{\catcode `\\12\catcode `\$12\catcode
  `\&12\catcode `\#12\catcode `\^12\catcode `\_12\catcode `\%12\relax}%
\providecommand \@@startlink[1]{}%
\providecommand \@@endlink[0]{}%
\providecommand \url  [0]{\begingroup\@sanitize@url \@url }%
\providecommand \@url [1]{\endgroup\@href {#1}{\urlprefix }}%
\providecommand \urlprefix  [0]{URL }%
\providecommand \Eprint [0]{\href }%
\providecommand \doibase [0]{http://dx.doi.org/}%
\providecommand \selectlanguage [0]{\@gobble}%
\providecommand \bibinfo  [0]{\@secondoftwo}%
\providecommand \bibfield  [0]{\@secondoftwo}%
\providecommand \translation [1]{[#1]}%
\providecommand \BibitemOpen [0]{}%
\providecommand \bibitemStop [0]{}%
\providecommand \bibitemNoStop [0]{.\EOS\space}%
\providecommand \EOS [0]{\spacefactor3000\relax}%
\providecommand \BibitemShut  [1]{\csname bibitem#1\endcsname}%
\let\auto@bib@innerbib\@empty
%</preamble>
\bibitem [{\citenamefont {van~de Krol}\ and\ \citenamefont
  {Parkinson}(2017)}]{Krol_perspective_photoelectrochemical_energy_storage_2017}%
  \BibitemOpen
  \bibfield  {author} {\bibinfo {author} {\bibfnamefont {R.}~\bibnamefont
  {van~de Krol}}\ and\ \bibinfo {author} {\bibfnamefont {B.~A.}\ \bibnamefont
  {Parkinson}},\ }\bibfield  {title} {\enquote {\bibinfo {title} {Perspectives
  on the photoelectrochemical storage of solar energy},}\ }\href {\doibase
  10.1557/mre.2017.15} {\bibfield  {journal} {\bibinfo  {journal} {MRS Energy
  Sust.}\ }\textbf {\bibinfo {volume} {4}},\ \bibinfo {pages} {e13} (\bibinfo
  {year} {2017})}\BibitemShut {NoStop}%
\bibitem [{\citenamefont {Ardo}\ \emph {et~al.}(2018)\citenamefont {Ardo},
  \citenamefont {Fernandez~Rivas}, \citenamefont {Modestino}, \citenamefont
  {Schulze~Greiving}, \citenamefont {Abdi}, \citenamefont {Alarcon-Llado},
  \citenamefont {Artero}, \citenamefont {Ayers}, \citenamefont {Battaglia},
  \citenamefont {Becker}, \citenamefont {Bederak}, \citenamefont {Berger},
  \citenamefont {Buda}, \citenamefont {Chinello}, \citenamefont {Dam},
  \citenamefont {Di~Palma}, \citenamefont {Edvinsson}, \citenamefont {Fujii},
  \citenamefont {Gardeniers}, \citenamefont {Geerlings}, \citenamefont
  {Hosseini~Hashemi}, \citenamefont {Haussener}, \citenamefont {Houle},
  \citenamefont {Huskens}, \citenamefont {James}, \citenamefont {Konrad},
  \citenamefont {Kudo}, \citenamefont {Kunturu}, \citenamefont {Lohse},
  \citenamefont {Mei}, \citenamefont {Miller}, \citenamefont {Moere},
  \citenamefont {Muller}, \citenamefont {Orchard}, \citenamefont {Rosser},
  \citenamefont {Saadi}, \citenamefont {Schuttauf}, \citenamefont {Seger},
  \citenamefont {Sheehan}, \citenamefont {Smith}, \citenamefont {Spurgeon},
  \citenamefont {Tang}, \citenamefont {van~de Krol}, \citenamefont {Vesborg},\
  and\ \citenamefont
  {Westerik}}]{Ardo_pathways_electrochemical_solar_hydrogen_2018}%
  \BibitemOpen
  \bibfield  {author} {\bibinfo {author} {\bibfnamefont {S.}~\bibnamefont
  {Ardo}}, \bibinfo {author} {\bibfnamefont {D.}~\bibnamefont
  {Fernandez~Rivas}}, \bibinfo {author} {\bibfnamefont {M.~A.}\ \bibnamefont
  {Modestino}}, \bibinfo {author} {\bibfnamefont {V.}~\bibnamefont
  {Schulze~Greiving}}, \bibinfo {author} {\bibfnamefont {F.}~\bibnamefont
  {Abdi}}, \bibinfo {author} {\bibfnamefont {E.}~\bibnamefont {Alarcon-Llado}},
  \bibinfo {author} {\bibfnamefont {V.}~\bibnamefont {Artero}}, \bibinfo
  {author} {\bibfnamefont {K.~E.}\ \bibnamefont {Ayers}}, \bibinfo {author}
  {\bibfnamefont {C.}~\bibnamefont {Battaglia}}, \bibinfo {author}
  {\bibfnamefont {J.-P.}\ \bibnamefont {Becker}}, \bibinfo {author}
  {\bibfnamefont {D.}~\bibnamefont {Bederak}}, \bibinfo {author} {\bibfnamefont
  {A.}~\bibnamefont {Berger}}, \bibinfo {author} {\bibfnamefont
  {F.}~\bibnamefont {Buda}}, \bibinfo {author} {\bibfnamefont {E.}~\bibnamefont
  {Chinello}}, \bibinfo {author} {\bibfnamefont {B.}~\bibnamefont {Dam}},
  \bibinfo {author} {\bibfnamefont {V.}~\bibnamefont {Di~Palma}}, \bibinfo
  {author} {\bibfnamefont {T.}~\bibnamefont {Edvinsson}}, \bibinfo {author}
  {\bibfnamefont {K.}~\bibnamefont {Fujii}}, \bibinfo {author} {\bibfnamefont
  {H.~J.}\ \bibnamefont {Gardeniers}}, \bibinfo {author} {\bibfnamefont
  {H.}~\bibnamefont {Geerlings}}, \bibinfo {author} {\bibfnamefont {S.~M.}\
  \bibnamefont {Hosseini~Hashemi}}, \bibinfo {author} {\bibfnamefont
  {S.}~\bibnamefont {Haussener}}, \bibinfo {author} {\bibfnamefont {F.~A.}\
  \bibnamefont {Houle}}, \bibinfo {author} {\bibfnamefont {J.}~\bibnamefont
  {Huskens}}, \bibinfo {author} {\bibfnamefont {B.}~\bibnamefont {James}},
  \bibinfo {author} {\bibfnamefont {K.}~\bibnamefont {Konrad}}, \bibinfo
  {author} {\bibfnamefont {A.}~\bibnamefont {Kudo}}, \bibinfo {author}
  {\bibfnamefont {P.~P.}\ \bibnamefont {Kunturu}}, \bibinfo {author}
  {\bibfnamefont {D.}~\bibnamefont {Lohse}}, \bibinfo {author} {\bibfnamefont
  {B.}~\bibnamefont {Mei}}, \bibinfo {author} {\bibfnamefont {E.}~\bibnamefont
  {Miller}}, \bibinfo {author} {\bibfnamefont {G.}~\bibnamefont {Moere}},
  \bibinfo {author} {\bibfnamefont {J.}~\bibnamefont {Muller}}, \bibinfo
  {author} {\bibfnamefont {K.~L.}\ \bibnamefont {Orchard}}, \bibinfo {author}
  {\bibfnamefont {T.}~\bibnamefont {Rosser}}, \bibinfo {author} {\bibfnamefont
  {F.~H.}\ \bibnamefont {Saadi}}, \bibinfo {author} {\bibfnamefont {J.-W.}\
  \bibnamefont {Schuttauf}}, \bibinfo {author} {\bibfnamefont {B.~J.}\
  \bibnamefont {Seger}}, \bibinfo {author} {\bibfnamefont {S.~W.}\ \bibnamefont
  {Sheehan}}, \bibinfo {author} {\bibfnamefont {W.~A.}\ \bibnamefont {Smith}},
  \bibinfo {author} {\bibfnamefont {J.}~\bibnamefont {Spurgeon}}, \bibinfo
  {author} {\bibfnamefont {M.}~\bibnamefont {Tang}}, \bibinfo {author}
  {\bibfnamefont {R.}~\bibnamefont {van~de Krol}}, \bibinfo {author}
  {\bibfnamefont {P.~C.~K.}\ \bibnamefont {Vesborg}}, \ and\ \bibinfo {author}
  {\bibfnamefont {P.}~\bibnamefont {Westerik}},\ }\bibfield  {title} {\enquote
  {\bibinfo {title} {Pathways to electrochemical solar-hydrogen
  technologies},}\ }\href {\doibase 10.1039/C7EE03639F} {\bibfield  {journal}
  {\bibinfo  {journal} {Energy Environ. Sci.}\ }\textbf {\bibinfo {volume}
  {11}},\ \bibinfo {pages} {2768--2783} (\bibinfo {year} {2018})}\BibitemShut
  {NoStop}%
\bibitem [{\citenamefont {Shaner}\ \emph {et~al.}(2016)\citenamefont {Shaner},
  \citenamefont {Atwater}, \citenamefont {Lewis},\ and\ \citenamefont
  {McFarland}}]{Shaner_technoeconomic_analysis_renwable_H2_2016}%
  \BibitemOpen
  \bibfield  {author} {\bibinfo {author} {\bibfnamefont {M.~R.}\ \bibnamefont
  {Shaner}}, \bibinfo {author} {\bibfnamefont {H.~A.}\ \bibnamefont {Atwater}},
  \bibinfo {author} {\bibfnamefont {N.~S.}\ \bibnamefont {Lewis}}, \ and\
  \bibinfo {author} {\bibfnamefont {E.~W.}\ \bibnamefont {McFarland}},\
  }\bibfield  {title} {\enquote {\bibinfo {title} {A comparative technoeconomic
  analysis of renewable hydrogen production using solar energy},}\ }\href
  {\doibase 10.1039/C5EE02573G} {\bibfield  {journal} {\bibinfo  {journal}
  {Energy Environ. Sci.}\ }\textbf {\bibinfo {volume} {9}},\ \bibinfo {pages}
  {2354--2371} (\bibinfo {year} {2016})}\BibitemShut {NoStop}%
\bibitem [{\citenamefont {Cheng}\ \emph {et~al.}(2018)\citenamefont {Cheng},
  \citenamefont {Richter}, \citenamefont {May}, \citenamefont {Ohlmann},
  \citenamefont {Lackner}, \citenamefont {Dimroth}, \citenamefont {Hannappel},
  \citenamefont {Atwater},\ and\ \citenamefont {Lewerenz}}]{Cheng_tandem_2018}%
  \BibitemOpen
  \bibfield  {author} {\bibinfo {author} {\bibfnamefont {W.-H.}\ \bibnamefont
  {Cheng}}, \bibinfo {author} {\bibfnamefont {M.~H.}\ \bibnamefont {Richter}},
  \bibinfo {author} {\bibfnamefont {M.~M.}\ \bibnamefont {May}}, \bibinfo
  {author} {\bibfnamefont {J.}~\bibnamefont {Ohlmann}}, \bibinfo {author}
  {\bibfnamefont {D.}~\bibnamefont {Lackner}}, \bibinfo {author} {\bibfnamefont
  {F.}~\bibnamefont {Dimroth}}, \bibinfo {author} {\bibfnamefont
  {T.}~\bibnamefont {Hannappel}}, \bibinfo {author} {\bibfnamefont {H.~A.}\
  \bibnamefont {Atwater}}, \ and\ \bibinfo {author} {\bibfnamefont {H.-J.}\
  \bibnamefont {Lewerenz}},\ }\bibfield  {title} {\enquote {\bibinfo {title}
  {Monolithic photoelectrochemical device for direct water splitting with 19\%
  efficiency},}\ }\href {\doibase 10.1021/acsenergylett.8b00920} {\bibfield
  {journal} {\bibinfo  {journal} {ACS Energy Lett.}\ }\textbf {\bibinfo
  {volume} {3}},\ \bibinfo {pages} {1795--1800} (\bibinfo {year}
  {2018})}\BibitemShut {NoStop}%
\bibitem [{\citenamefont {Pinaud}\ \emph {et~al.}(2013)\citenamefont {Pinaud},
  \citenamefont {Benck}, \citenamefont {Seitz}, \citenamefont {Forman},
  \citenamefont {Chen}, \citenamefont {Deutsch}, \citenamefont {James},
  \citenamefont {Baum}, \citenamefont {Baum}, \citenamefont {Ardo},
  \citenamefont {Wang}, \citenamefont {Miller},\ and\ \citenamefont
  {Jaramillo}}]{Pinaud_hydrogen_production_facilities_2013}%
  \BibitemOpen
  \bibfield  {author} {\bibinfo {author} {\bibfnamefont {B.~A.}\ \bibnamefont
  {Pinaud}}, \bibinfo {author} {\bibfnamefont {J.~D.}\ \bibnamefont {Benck}},
  \bibinfo {author} {\bibfnamefont {L.~C.}\ \bibnamefont {Seitz}}, \bibinfo
  {author} {\bibfnamefont {A.~J.}\ \bibnamefont {Forman}}, \bibinfo {author}
  {\bibfnamefont {Z.}~\bibnamefont {Chen}}, \bibinfo {author} {\bibfnamefont
  {T.~G.}\ \bibnamefont {Deutsch}}, \bibinfo {author} {\bibfnamefont {B.~D.}\
  \bibnamefont {James}}, \bibinfo {author} {\bibfnamefont {K.~N.}\ \bibnamefont
  {Baum}}, \bibinfo {author} {\bibfnamefont {G.~N.}\ \bibnamefont {Baum}},
  \bibinfo {author} {\bibfnamefont {S.}~\bibnamefont {Ardo}}, \bibinfo {author}
  {\bibfnamefont {H.}~\bibnamefont {Wang}}, \bibinfo {author} {\bibfnamefont
  {E.}~\bibnamefont {Miller}}, \ and\ \bibinfo {author} {\bibfnamefont {T.~F.}\
  \bibnamefont {Jaramillo}},\ }\bibfield  {title} {\enquote {\bibinfo {title}
  {{Technical and economic feasibility of centralized facilities for solar
  hydrogen production via photocatalysis and photoelectrochemistry}},}\ }\href
  {\doibase 10.1039/C3EE40831K} {\bibfield  {journal} {\bibinfo  {journal}
  {Energy Environ. Sci.}\ }\textbf {\bibinfo {volume} {6}},\ \bibinfo {pages}
  {1983--2002} (\bibinfo {year} {2013})}\BibitemShut {NoStop}%
\bibitem [{\citenamefont {Tournet}\ \emph {et~al.}(2020)\citenamefont
  {Tournet}, \citenamefont {Lee}, \citenamefont {Karuturi}, \citenamefont
  {Tan},\ and\ \citenamefont
  {Jagadish}}]{Tournet_review_III-V_water_splitting_2020}%
  \BibitemOpen
  \bibfield  {author} {\bibinfo {author} {\bibfnamefont {J.}~\bibnamefont
  {Tournet}}, \bibinfo {author} {\bibfnamefont {Y.}~\bibnamefont {Lee}},
  \bibinfo {author} {\bibfnamefont {S.~K.}\ \bibnamefont {Karuturi}}, \bibinfo
  {author} {\bibfnamefont {H.~H.}\ \bibnamefont {Tan}}, \ and\ \bibinfo
  {author} {\bibfnamefont {C.}~\bibnamefont {Jagadish}},\ }\bibfield  {title}
  {\enquote {\bibinfo {title} {Iii-v semiconductor materials for solar hydrogen
  production: Status and prospects},}\ }\href {\doibase
  10.1021/acsenergylett.9b02582} {\bibfield  {journal} {\bibinfo  {journal}
  {ACS Energy Lett.}\ }\textbf {\bibinfo {volume} {5}},\ \bibinfo {pages}
  {611--622} (\bibinfo {year} {2020})}\BibitemShut {NoStop}%
\bibitem [{\citenamefont {Cariou}\ \emph {et~al.}(2018)\citenamefont {Cariou},
  \citenamefont {Benick}, \citenamefont {Feldmann}, \citenamefont {H{\"o}hn},
  \citenamefont {Hauser}, \citenamefont {Beutel}, \citenamefont {Razek},
  \citenamefont {Wimplinger}, \citenamefont {Bl{\"a}si}, \citenamefont
  {Lackner}, \citenamefont {Hermle}, \citenamefont {Siefer}, \citenamefont
  {Glunz}, \citenamefont {Bett},\ and\ \citenamefont {Dimroth}}]{Cariou2018}%
  \BibitemOpen
  \bibfield  {author} {\bibinfo {author} {\bibfnamefont {R.}~\bibnamefont
  {Cariou}}, \bibinfo {author} {\bibfnamefont {J.}~\bibnamefont {Benick}},
  \bibinfo {author} {\bibfnamefont {F.}~\bibnamefont {Feldmann}}, \bibinfo
  {author} {\bibfnamefont {O.}~\bibnamefont {H{\"o}hn}}, \bibinfo {author}
  {\bibfnamefont {H.}~\bibnamefont {Hauser}}, \bibinfo {author} {\bibfnamefont
  {P.}~\bibnamefont {Beutel}}, \bibinfo {author} {\bibfnamefont
  {N.}~\bibnamefont {Razek}}, \bibinfo {author} {\bibfnamefont
  {M.}~\bibnamefont {Wimplinger}}, \bibinfo {author} {\bibfnamefont
  {B.}~\bibnamefont {Bl{\"a}si}}, \bibinfo {author} {\bibfnamefont
  {D.}~\bibnamefont {Lackner}}, \bibinfo {author} {\bibfnamefont
  {M.}~\bibnamefont {Hermle}}, \bibinfo {author} {\bibfnamefont
  {G.}~\bibnamefont {Siefer}}, \bibinfo {author} {\bibfnamefont {S.~W.}\
  \bibnamefont {Glunz}}, \bibinfo {author} {\bibfnamefont {A.~W.}\ \bibnamefont
  {Bett}}, \ and\ \bibinfo {author} {\bibfnamefont {F.}~\bibnamefont
  {Dimroth}},\ }\bibfield  {title} {\enquote {\bibinfo {title}
  {{III--V-on-silicon solar cells reaching 33{\%} photoconversion efficiency in
  two-terminal configuration}},}\ }\href {\doibase 10.1038/s41560-018-0125-0}
  {\bibfield  {journal} {\bibinfo  {journal} {Nat. Energy}\ }\textbf {\bibinfo
  {volume} {3}},\ \bibinfo {pages} {326--333} (\bibinfo {year}
  {2018})}\BibitemShut {NoStop}%
\bibitem [{\citenamefont {Chen}\ \emph {et~al.}(2021)\citenamefont {Chen},
  \citenamefont {Alqahtani}, \citenamefont {Levallois}, \citenamefont
  {Létoublon}, \citenamefont {Stervinou}, \citenamefont {Piron}, \citenamefont
  {Boyer-Richard}, \citenamefont {Jancu}, \citenamefont {Rohel}, \citenamefont
  {Bernard}, \citenamefont {Léger}, \citenamefont {Bertru}, \citenamefont
  {Wu}, \citenamefont {Parkin},\ and\ \citenamefont
  {Cornet}}]{Chen_assessment_GaPSb-Si_photoelectrochemical_cells_2021}%
  \BibitemOpen
  \bibfield  {author} {\bibinfo {author} {\bibfnamefont {L.}~\bibnamefont
  {Chen}}, \bibinfo {author} {\bibfnamefont {M.}~\bibnamefont {Alqahtani}},
  \bibinfo {author} {\bibfnamefont {C.}~\bibnamefont {Levallois}}, \bibinfo
  {author} {\bibfnamefont {A.}~\bibnamefont {Létoublon}}, \bibinfo {author}
  {\bibfnamefont {J.}~\bibnamefont {Stervinou}}, \bibinfo {author}
  {\bibfnamefont {R.}~\bibnamefont {Piron}}, \bibinfo {author} {\bibfnamefont
  {S.}~\bibnamefont {Boyer-Richard}}, \bibinfo {author} {\bibfnamefont {J.-M.}\
  \bibnamefont {Jancu}}, \bibinfo {author} {\bibfnamefont {T.}~\bibnamefont
  {Rohel}}, \bibinfo {author} {\bibfnamefont {R.}~\bibnamefont {Bernard}},
  \bibinfo {author} {\bibfnamefont {Y.}~\bibnamefont {Léger}}, \bibinfo
  {author} {\bibfnamefont {N.}~\bibnamefont {Bertru}}, \bibinfo {author}
  {\bibfnamefont {J.}~\bibnamefont {Wu}}, \bibinfo {author} {\bibfnamefont
  {I.~P.}\ \bibnamefont {Parkin}}, \ and\ \bibinfo {author} {\bibfnamefont
  {C.}~\bibnamefont {Cornet}},\ }\bibfield  {title} {\enquote {\bibinfo {title}
  {{Assessment of GaPSb/Si tandem material association properties for
  photoelectrochemical cells}},}\ }\href {\doibase
  10.1016/j.solmat.2020.110888} {\bibfield  {journal} {\bibinfo  {journal}
  {Solar Energy Materials and Solar Cells}\ }\textbf {\bibinfo {volume}
  {221}},\ \bibinfo {pages} {110888} (\bibinfo {year} {2021})}\BibitemShut
  {NoStop}%
\bibitem [{\citenamefont {Vanka}\ \emph {et~al.}(2020)\citenamefont {Vanka},
  \citenamefont {Zhou}, \citenamefont {Awni}, \citenamefont {Song},
  \citenamefont {Chowdhury}, \citenamefont {Liu}, \citenamefont {Hajibabaei},
  \citenamefont {Shi}, \citenamefont {Xiao}, \citenamefont {Navid},
  \citenamefont {Pandey}, \citenamefont {Chen}, \citenamefont {Botton},
  \citenamefont {Hamann}, \citenamefont {Wang}, \citenamefont {Yan},\ and\
  \citenamefont {Mi}}]{Vanka_InGaN-Si_water_splitting_2020}%
  \BibitemOpen
  \bibfield  {author} {\bibinfo {author} {\bibfnamefont {S.}~\bibnamefont
  {Vanka}}, \bibinfo {author} {\bibfnamefont {B.}~\bibnamefont {Zhou}},
  \bibinfo {author} {\bibfnamefont {R.~A.}\ \bibnamefont {Awni}}, \bibinfo
  {author} {\bibfnamefont {Z.}~\bibnamefont {Song}}, \bibinfo {author}
  {\bibfnamefont {F.~A.}\ \bibnamefont {Chowdhury}}, \bibinfo {author}
  {\bibfnamefont {X.}~\bibnamefont {Liu}}, \bibinfo {author} {\bibfnamefont
  {H.}~\bibnamefont {Hajibabaei}}, \bibinfo {author} {\bibfnamefont
  {W.}~\bibnamefont {Shi}}, \bibinfo {author} {\bibfnamefont {Y.}~\bibnamefont
  {Xiao}}, \bibinfo {author} {\bibfnamefont {I.~A.}\ \bibnamefont {Navid}},
  \bibinfo {author} {\bibfnamefont {A.}~\bibnamefont {Pandey}}, \bibinfo
  {author} {\bibfnamefont {R.}~\bibnamefont {Chen}}, \bibinfo {author}
  {\bibfnamefont {G.~A.}\ \bibnamefont {Botton}}, \bibinfo {author}
  {\bibfnamefont {T.~W.}\ \bibnamefont {Hamann}}, \bibinfo {author}
  {\bibfnamefont {D.}~\bibnamefont {Wang}}, \bibinfo {author} {\bibfnamefont
  {Y.}~\bibnamefont {Yan}}, \ and\ \bibinfo {author} {\bibfnamefont
  {Z.}~\bibnamefont {Mi}},\ }\bibfield  {title} {\enquote {\bibinfo {title}
  {{InGaN/Si Double-Junction Photocathode for Unassisted Solar Water
  Splitting}},}\ }\href {\doibase 10.1021/acsenergylett.0c01583} {\bibfield
  {journal} {\bibinfo  {journal} {ACS Energy Lett.}\ }\textbf {\bibinfo
  {volume} {5}},\ \bibinfo {pages} {3741--3751} (\bibinfo {year}
  {2020})}\BibitemShut {NoStop}%
\bibitem [{\citenamefont {Lackner}\ \emph {et~al.}(2020)\citenamefont
  {Lackner}, \citenamefont {H{\"o}hn}, \citenamefont {M{\"u}ller},
  \citenamefont {Beutel}, \citenamefont {Schygulla}, \citenamefont {Hauser},
  \citenamefont {Predan}, \citenamefont {Siefer}, \citenamefont {Schachtner},
  \citenamefont {Sch{\"o}n}, \citenamefont {Benick}, \citenamefont {Hermle},\
  and\ \citenamefont {Dimroth}}]{Lackner_2020}%
  \BibitemOpen
  \bibfield  {author} {\bibinfo {author} {\bibfnamefont {D.}~\bibnamefont
  {Lackner}}, \bibinfo {author} {\bibfnamefont {O.}~\bibnamefont {H{\"o}hn}},
  \bibinfo {author} {\bibfnamefont {R.}~\bibnamefont {M{\"u}ller}}, \bibinfo
  {author} {\bibfnamefont {P.}~\bibnamefont {Beutel}}, \bibinfo {author}
  {\bibfnamefont {P.}~\bibnamefont {Schygulla}}, \bibinfo {author}
  {\bibfnamefont {H.}~\bibnamefont {Hauser}}, \bibinfo {author} {\bibfnamefont
  {F.}~\bibnamefont {Predan}}, \bibinfo {author} {\bibfnamefont
  {G.}~\bibnamefont {Siefer}}, \bibinfo {author} {\bibfnamefont
  {M.}~\bibnamefont {Schachtner}}, \bibinfo {author} {\bibfnamefont
  {J.}~\bibnamefont {Sch{\"o}n}}, \bibinfo {author} {\bibfnamefont
  {J.}~\bibnamefont {Benick}}, \bibinfo {author} {\bibfnamefont
  {M.}~\bibnamefont {Hermle}}, \ and\ \bibinfo {author} {\bibfnamefont
  {F.}~\bibnamefont {Dimroth}},\ }\bibfield  {title} {\enquote {\bibinfo
  {title} {Two-terminal direct wafer-bonded {GaInP/AlGaAs//Si} triple-junction
  solar cell with {AM}1.5g efficiency of 34.1\%},}\ }\href {\doibase
  10.1002/solr.202000210} {\bibfield  {journal} {\bibinfo  {journal} {Solar
  RRL}\ }\textbf {\bibinfo {volume} {4}},\ \bibinfo {pages} {2000210} (\bibinfo
  {year} {2020})}\BibitemShut {NoStop}%
\bibitem [{ISE()}]{ISE_press_release}%
  \BibitemOpen
  \href@noop {} {\enquote {\bibinfo {title} {Fraunhofer {ISE} press release.
  https://www.ise.fraunhofer.de/en/press-media/press-releases/2021/tandem-photovoltaics-enables-new-heights-in-solar-cell-efficiencies-35point9-percent-for-iii-v-silicon-solar-cell.html
  (accessed on 15.06.2021)},}\ }\BibitemShut {NoStop}%
\bibitem [{\citenamefont {Feifel}\ \emph {et~al.}(2019)\citenamefont {Feifel},
  \citenamefont {Lackner}, \citenamefont {Ohlmann}, \citenamefont {Benick},
  \citenamefont {Hermle},\ and\ \citenamefont
  {Dimroth}}]{Feifel_directly_grown_III-V_Si_cell_2019}%
  \BibitemOpen
  \bibfield  {author} {\bibinfo {author} {\bibfnamefont {M.}~\bibnamefont
  {Feifel}}, \bibinfo {author} {\bibfnamefont {D.}~\bibnamefont {Lackner}},
  \bibinfo {author} {\bibfnamefont {J.}~\bibnamefont {Ohlmann}}, \bibinfo
  {author} {\bibfnamefont {J.}~\bibnamefont {Benick}}, \bibinfo {author}
  {\bibfnamefont {M.}~\bibnamefont {Hermle}}, \ and\ \bibinfo {author}
  {\bibfnamefont {F.}~\bibnamefont {Dimroth}},\ }\bibfield  {title} {\enquote
  {\bibinfo {title} {{Direct Growth of a GaInP/GaAs/Si Triple-Junction Solar
  Cell with 22.3\% AM1.5g Efficiency}},}\ }\href {\doibase
  10.1002/solr.201900313} {\bibfield  {journal} {\bibinfo  {journal} {Solar
  RRL}\ }\textbf {\bibinfo {volume} {3}},\ \bibinfo {pages} {1900313} (\bibinfo
  {year} {2019})}\BibitemShut {NoStop}%
\bibitem [{\citenamefont {Feifel}\ \emph {et~al.}(2021)\citenamefont {Feifel},
  \citenamefont {Lackner}, \citenamefont {Sch{\"o}n}, \citenamefont {Ohlmann},
  \citenamefont {Benick}, \citenamefont {Siefer}, \citenamefont {Predan},
  \citenamefont {Hermle},\ and\ \citenamefont {Dimroth}}]{Feifel_2021}%
  \BibitemOpen
  \bibfield  {author} {\bibinfo {author} {\bibfnamefont {M.}~\bibnamefont
  {Feifel}}, \bibinfo {author} {\bibfnamefont {D.}~\bibnamefont {Lackner}},
  \bibinfo {author} {\bibfnamefont {J.}~\bibnamefont {Sch{\"o}n}}, \bibinfo
  {author} {\bibfnamefont {J.}~\bibnamefont {Ohlmann}}, \bibinfo {author}
  {\bibfnamefont {J.}~\bibnamefont {Benick}}, \bibinfo {author} {\bibfnamefont
  {G.}~\bibnamefont {Siefer}}, \bibinfo {author} {\bibfnamefont
  {F.}~\bibnamefont {Predan}}, \bibinfo {author} {\bibfnamefont
  {M.}~\bibnamefont {Hermle}}, \ and\ \bibinfo {author} {\bibfnamefont
  {F.}~\bibnamefont {Dimroth}},\ }\bibfield  {title} {\enquote {\bibinfo
  {title} {Epitaxial {GaInP/GaAs/Si} triple-junction solar cell with 25.9\%
  {AM}1.5g efficiency enabled by transparent metamorphic
  {Al$_{x}$Ga$_{1-x}$As$_{y}$P$_{1-y}$} step-graded buffer structures},}\
  }\href {\doibase 10.1002/solr.202000763} {\bibfield  {journal} {\bibinfo
  {journal} {Solar RRL}\ }\textbf {\bibinfo {volume} {5}},\ \bibinfo {pages}
  {2000763} (\bibinfo {year} {2021})}\BibitemShut {NoStop}%
\bibitem [{\citenamefont {Al-Ashouri}\ \emph {et~al.}(2020)\citenamefont
  {Al-Ashouri}, \citenamefont {K{\"o}hnen}, \citenamefont {Li}, \citenamefont
  {Magomedov}, \citenamefont {Hempel}, \citenamefont {Caprioglio},
  \citenamefont {M{\'a}rquez}, \citenamefont {Morales~Vilches}, \citenamefont
  {Kasparavicius}, \citenamefont {Smith}, \citenamefont {Phung}, \citenamefont
  {Menzel}, \citenamefont {Grischek}, \citenamefont {Kegelmann}, \citenamefont
  {Skroblin}, \citenamefont {Gollwitzer}, \citenamefont {Malinauskas},
  \citenamefont {Jo{\v s}t}, \citenamefont {Mati{\v c}}, \citenamefont {Rech},
  \citenamefont {Schlatmann}, \citenamefont {Topi{\v c}}, \citenamefont
  {Korte}, \citenamefont {Abate}, \citenamefont {Stannowski}, \citenamefont
  {Neher}, \citenamefont {Stolterfoht}, \citenamefont {Unold}, \citenamefont
  {Getautis},\ and\ \citenamefont {Albrecht}}]{Al-Ashouri1300}%
  \BibitemOpen
  \bibfield  {author} {\bibinfo {author} {\bibfnamefont {A.}~\bibnamefont
  {Al-Ashouri}}, \bibinfo {author} {\bibfnamefont {E.}~\bibnamefont
  {K{\"o}hnen}}, \bibinfo {author} {\bibfnamefont {B.}~\bibnamefont {Li}},
  \bibinfo {author} {\bibfnamefont {A.}~\bibnamefont {Magomedov}}, \bibinfo
  {author} {\bibfnamefont {H.}~\bibnamefont {Hempel}}, \bibinfo {author}
  {\bibfnamefont {P.}~\bibnamefont {Caprioglio}}, \bibinfo {author}
  {\bibfnamefont {J.~A.}\ \bibnamefont {M{\'a}rquez}}, \bibinfo {author}
  {\bibfnamefont {A.~B.}\ \bibnamefont {Morales~Vilches}}, \bibinfo {author}
  {\bibfnamefont {E.}~\bibnamefont {Kasparavicius}}, \bibinfo {author}
  {\bibfnamefont {J.~A.}\ \bibnamefont {Smith}}, \bibinfo {author}
  {\bibfnamefont {N.}~\bibnamefont {Phung}}, \bibinfo {author} {\bibfnamefont
  {D.}~\bibnamefont {Menzel}}, \bibinfo {author} {\bibfnamefont
  {M.}~\bibnamefont {Grischek}}, \bibinfo {author} {\bibfnamefont
  {L.}~\bibnamefont {Kegelmann}}, \bibinfo {author} {\bibfnamefont
  {D.}~\bibnamefont {Skroblin}}, \bibinfo {author} {\bibfnamefont
  {C.}~\bibnamefont {Gollwitzer}}, \bibinfo {author} {\bibfnamefont
  {T.}~\bibnamefont {Malinauskas}}, \bibinfo {author} {\bibfnamefont
  {M.}~\bibnamefont {Jo{\v s}t}}, \bibinfo {author} {\bibfnamefont
  {G.}~\bibnamefont {Mati{\v c}}}, \bibinfo {author} {\bibfnamefont
  {B.}~\bibnamefont {Rech}}, \bibinfo {author} {\bibfnamefont {R.}~\bibnamefont
  {Schlatmann}}, \bibinfo {author} {\bibfnamefont {M.}~\bibnamefont {Topi{\v
  c}}}, \bibinfo {author} {\bibfnamefont {L.}~\bibnamefont {Korte}}, \bibinfo
  {author} {\bibfnamefont {A.}~\bibnamefont {Abate}}, \bibinfo {author}
  {\bibfnamefont {B.}~\bibnamefont {Stannowski}}, \bibinfo {author}
  {\bibfnamefont {D.}~\bibnamefont {Neher}}, \bibinfo {author} {\bibfnamefont
  {M.}~\bibnamefont {Stolterfoht}}, \bibinfo {author} {\bibfnamefont
  {T.}~\bibnamefont {Unold}}, \bibinfo {author} {\bibfnamefont
  {V.}~\bibnamefont {Getautis}}, \ and\ \bibinfo {author} {\bibfnamefont
  {S.}~\bibnamefont {Albrecht}},\ }\bibfield  {title} {\enquote {\bibinfo
  {title} {Monolithic perovskite/silicon tandem solar cell with >29\%
  efficiency by enhanced hole extraction},}\ }\href {\doibase
  10.1126/science.abd4016} {\bibfield  {journal} {\bibinfo  {journal}
  {Science}\ }\textbf {\bibinfo {volume} {370}},\ \bibinfo {pages} {1300--1309}
  (\bibinfo {year} {2020})}\BibitemShut {NoStop}%
\bibitem [{\citenamefont {Köhnen}\ \emph {et~al.}(2021)\citenamefont
  {Köhnen}, \citenamefont {Wagner}, \citenamefont {Lang}, \citenamefont
  {Cruz}, \citenamefont {Li}, \citenamefont {Roß}, \citenamefont {Jošt},
  \citenamefont {Morales-Vilches}, \citenamefont {Topič}, \citenamefont
  {Stolterfoht}, \citenamefont {Neher}, \citenamefont {Korte}, \citenamefont
  {Rech}, \citenamefont {Schlatmann}, \citenamefont {Stannowski},\ and\
  \citenamefont {Albrecht}}]{Koehnen_2021}%
  \BibitemOpen
  \bibfield  {author} {\bibinfo {author} {\bibfnamefont {E.}~\bibnamefont
  {Köhnen}}, \bibinfo {author} {\bibfnamefont {P.}~\bibnamefont {Wagner}},
  \bibinfo {author} {\bibfnamefont {F.}~\bibnamefont {Lang}}, \bibinfo {author}
  {\bibfnamefont {A.}~\bibnamefont {Cruz}}, \bibinfo {author} {\bibfnamefont
  {B.}~\bibnamefont {Li}}, \bibinfo {author} {\bibfnamefont {M.}~\bibnamefont
  {Roß}}, \bibinfo {author} {\bibfnamefont {M.}~\bibnamefont {Jošt}},
  \bibinfo {author} {\bibfnamefont {A.~B.}\ \bibnamefont {Morales-Vilches}},
  \bibinfo {author} {\bibfnamefont {M.}~\bibnamefont {Topič}}, \bibinfo
  {author} {\bibfnamefont {M.}~\bibnamefont {Stolterfoht}}, \bibinfo {author}
  {\bibfnamefont {D.}~\bibnamefont {Neher}}, \bibinfo {author} {\bibfnamefont
  {L.}~\bibnamefont {Korte}}, \bibinfo {author} {\bibfnamefont
  {B.}~\bibnamefont {Rech}}, \bibinfo {author} {\bibfnamefont {R.}~\bibnamefont
  {Schlatmann}}, \bibinfo {author} {\bibfnamefont {B.}~\bibnamefont
  {Stannowski}}, \ and\ \bibinfo {author} {\bibfnamefont {S.}~\bibnamefont
  {Albrecht}},\ }\bibfield  {title} {\enquote {\bibinfo {title} {27.9\%
  efficient monolithic perovskite/silicon tandem solar cells on industry
  compatible bottom cells},}\ }\href {\doibase 10.1002/solr.202100244}
  {\bibfield  {journal} {\bibinfo  {journal} {Solar RRL}\ }\textbf {\bibinfo
  {volume} {n/a}},\ \bibinfo {pages} {2100244} (\bibinfo {year}
  {2021})}\BibitemShut {NoStop}%
\bibitem [{\citenamefont {May}, \citenamefont {Döscher},\ and\ \citenamefont
  {Turner}(2018)}]{High_efficiency_book_chapter_2018}%
  \BibitemOpen
  \bibfield  {author} {\bibinfo {author} {\bibfnamefont {M.~M.}\ \bibnamefont
  {May}}, \bibinfo {author} {\bibfnamefont {H.}~\bibnamefont {Döscher}}, \
  and\ \bibinfo {author} {\bibfnamefont {J.}~\bibnamefont {Turner}},\
  }\bibfield  {title} {\enquote {\bibinfo {title} {{High-efficiency water
  splitting systems}},}\ }in\ \href {\doibase 10.1039/9781788010313-00454}
  {\emph {\bibinfo {booktitle} {{Integrated Solar Fuel Generators}}}},\
  \bibinfo {series and number} {Energy and Environment Series},\ \bibinfo
  {editor} {edited by\ \bibinfo {editor} {\bibfnamefont {I.~D.}\ \bibnamefont
  {Sharp}}, \bibinfo {editor} {\bibfnamefont {H.~A.}\ \bibnamefont {Atwater}},
  \ and\ \bibinfo {editor} {\bibfnamefont {H.-J.}\ \bibnamefont {Lewerenz}}}\
  (\bibinfo  {publisher} {The Royal Society of Chemistry},\ \bibinfo {year}
  {2018})\ Chap.~\bibinfo {chapter} {12}, pp.\ \bibinfo {pages}
  {454--499}\BibitemShut {NoStop}%
\bibitem [{\citenamefont
  {Parkinson}(1984)}]{Parkinson_efficiency_stability_PEC_devices_1984}%
  \BibitemOpen
  \bibfield  {author} {\bibinfo {author} {\bibfnamefont {B.}~\bibnamefont
  {Parkinson}},\ }\bibfield  {title} {\enquote {\bibinfo {title} {On the
  efficiency and stability of photoelectrochemical devices},}\ }\href {\doibase
  10.1021/ar00108a004} {\bibfield  {journal} {\bibinfo  {journal} {Acc. Chem.
  Res.}\ }\textbf {\bibinfo {volume} {17}},\ \bibinfo {pages} {431--437}
  (\bibinfo {year} {1984})}\BibitemShut {NoStop}%
\bibitem [{\citenamefont {Murphy}\ \emph {et~al.}(2006)\citenamefont {Murphy},
  \citenamefont {Barnes}, \citenamefont {Randeniya}, \citenamefont {Plumb},
  \citenamefont {Grey}, \citenamefont {Horne},\ and\ \citenamefont
  {Glasscock}}]{Murphy_water_filter_water_splitting_2006}%
  \BibitemOpen
  \bibfield  {author} {\bibinfo {author} {\bibfnamefont {A.}~\bibnamefont
  {Murphy}}, \bibinfo {author} {\bibfnamefont {P.}~\bibnamefont {Barnes}},
  \bibinfo {author} {\bibfnamefont {L.}~\bibnamefont {Randeniya}}, \bibinfo
  {author} {\bibfnamefont {I.}~\bibnamefont {Plumb}}, \bibinfo {author}
  {\bibfnamefont {I.}~\bibnamefont {Grey}}, \bibinfo {author} {\bibfnamefont
  {M.}~\bibnamefont {Horne}}, \ and\ \bibinfo {author} {\bibfnamefont
  {J.}~\bibnamefont {Glasscock}},\ }\bibfield  {title} {\enquote {\bibinfo
  {title} {Efficiency of solar water splitting using semiconductor
  electrodes},}\ }\href {\doibase 10.1016/j.ijhydene.2006.01.014} {\bibfield
  {journal} {\bibinfo  {journal} {Int. J. Hydrogen Energy}\ }\textbf {\bibinfo
  {volume} {31}},\ \bibinfo {pages} {1999 -- 2017} (\bibinfo {year}
  {2006})}\BibitemShut {NoStop}%
\bibitem [{\citenamefont {Döscher}\ \emph {et~al.}(2014)\citenamefont
  {Döscher}, \citenamefont {Geisz}, \citenamefont {Deutsch},\ and\
  \citenamefont {Turner}}]{Doescher_sunlight_water_2014}%
  \BibitemOpen
  \bibfield  {author} {\bibinfo {author} {\bibfnamefont {H.}~\bibnamefont
  {Döscher}}, \bibinfo {author} {\bibfnamefont {J.~F.}\ \bibnamefont {Geisz}},
  \bibinfo {author} {\bibfnamefont {T.~G.}\ \bibnamefont {Deutsch}}, \ and\
  \bibinfo {author} {\bibfnamefont {J.~A.}\ \bibnamefont {Turner}},\ }\bibfield
   {title} {\enquote {\bibinfo {title} {Sunlight absorption in water –
  efficiency and design implications for photoelectrochemical devices},}\
  }\href {\doibase 10.1039/C4EE01753F} {\bibfield  {journal} {\bibinfo
  {journal} {Energy Environ. Sci.}\ }\textbf {\bibinfo {volume} {7}},\ \bibinfo
  {pages} {2951--2956} (\bibinfo {year} {2014})}\BibitemShut {NoStop}%
\bibitem [{\citenamefont {Pope}\ and\ \citenamefont
  {Fry}(1997)}]{Pope_absorption_water_visible_1997}%
  \BibitemOpen
  \bibfield  {author} {\bibinfo {author} {\bibfnamefont {R.~M.}\ \bibnamefont
  {Pope}}\ and\ \bibinfo {author} {\bibfnamefont {E.~S.}\ \bibnamefont {Fry}},\
  }\bibfield  {title} {\enquote {\bibinfo {title} {Absorption spectrum
  (380--700 nm) of pure water. ii. integrating cavity measurements},}\ }\href
  {\doibase 10.1364/AO.36.008710} {\bibfield  {journal} {\bibinfo  {journal}
  {Appl. Opt.}\ }\textbf {\bibinfo {volume} {36}},\ \bibinfo {pages}
  {8710--8723} (\bibinfo {year} {1997})}\BibitemShut {NoStop}%
\bibitem [{\citenamefont {May}\ \emph {et~al.}(2017)\citenamefont {May},
  \citenamefont {Lackner}, \citenamefont {Ohlmann}, \citenamefont {Dimroth},
  \citenamefont {van~de Krol}, \citenamefont {Hannappel},\ and\ \citenamefont
  {Schwarzburg}}]{May_benchmarking_PEC_2017}%
  \BibitemOpen
  \bibfield  {author} {\bibinfo {author} {\bibfnamefont {M.~M.}\ \bibnamefont
  {May}}, \bibinfo {author} {\bibfnamefont {D.}~\bibnamefont {Lackner}},
  \bibinfo {author} {\bibfnamefont {J.}~\bibnamefont {Ohlmann}}, \bibinfo
  {author} {\bibfnamefont {F.}~\bibnamefont {Dimroth}}, \bibinfo {author}
  {\bibfnamefont {R.}~\bibnamefont {van~de Krol}}, \bibinfo {author}
  {\bibfnamefont {T.}~\bibnamefont {Hannappel}}, \ and\ \bibinfo {author}
  {\bibfnamefont {K.}~\bibnamefont {Schwarzburg}},\ }\bibfield  {title}
  {\enquote {\bibinfo {title} {On the benchmarking of multi-junction
  photoelectrochemical fuel generating devices},}\ }\href {\doibase
  10.1039/C6SE00083E} {\bibfield  {journal} {\bibinfo  {journal} {Sust. En.
  Fuels}\ }\textbf {\bibinfo {volume} {1}},\ \bibinfo {pages} {492--503}
  (\bibinfo {year} {2017})}\BibitemShut {NoStop}%
\bibitem [{\citenamefont {Gueymard}(2004)}]{Gueymard_2004}%
  \BibitemOpen
  \bibfield  {author} {\bibinfo {author} {\bibfnamefont {C.~A.}\ \bibnamefont
  {Gueymard}},\ }\bibfield  {title} {\enquote {\bibinfo {title} {The sun’s
  total and spectral irradiance for solar energy applications and solar
  radiation models},}\ }\href {\doibase
  https://doi.org/10.1016/j.solener.2003.08.039} {\bibfield  {journal}
  {\bibinfo  {journal} {Sol. Energy}\ }\textbf {\bibinfo {volume} {76}},\
  \bibinfo {pages} {423--453} (\bibinfo {year} {2004})}\BibitemShut {NoStop}%
\bibitem [{\citenamefont {Kou}, \citenamefont {Labrie},\ and\ \citenamefont
  {Chylek}(1993)}]{Kou_1993}%
  \BibitemOpen
  \bibfield  {author} {\bibinfo {author} {\bibfnamefont {L.}~\bibnamefont
  {Kou}}, \bibinfo {author} {\bibfnamefont {D.}~\bibnamefont {Labrie}}, \ and\
  \bibinfo {author} {\bibfnamefont {P.}~\bibnamefont {Chylek}},\ }\bibfield
  {title} {\enquote {\bibinfo {title} {Refractive indices of water and ice in
  the 0.65- to 2.5-$\mu$m spectral range},}\ }\href {\doibase
  10.1364/AO.32.003531} {\bibfield  {journal} {\bibinfo  {journal} {Appl.
  Opt.}\ }\textbf {\bibinfo {volume} {32}},\ \bibinfo {pages} {3531--3540}
  (\bibinfo {year} {1993})}\BibitemShut {NoStop}%
\bibitem [{\citenamefont {May}\ and\ \citenamefont
  {Kölbach}(2021)}]{May_yasofo-data_2020}%
  \BibitemOpen
  \bibfield  {author} {\bibinfo {author} {\bibfnamefont {M.~M.}\ \bibnamefont
  {May}}\ and\ \bibinfo {author} {\bibfnamefont {M.}~\bibnamefont {Kölbach}},\
  }\href {\doibase 10.5281/zenodo.1489157} {\enquote {\bibinfo {title} {{YaSoFo
  - Yet Another SOlar Fuels Optimizer (updated code to be published alongside
  the manuscript)}},}\ } (\bibinfo {year} {2021})\BibitemShut {NoStop}%
\bibitem [{\citenamefont {Obata}, \citenamefont {Mokeddem},\ and\ \citenamefont
  {Abdi}(2021)}]{Obata_bubbles_2021}%
  \BibitemOpen
  \bibfield  {author} {\bibinfo {author} {\bibfnamefont {K.}~\bibnamefont
  {Obata}}, \bibinfo {author} {\bibfnamefont {A.}~\bibnamefont {Mokeddem}}, \
  and\ \bibinfo {author} {\bibfnamefont {F.~F.}\ \bibnamefont {Abdi}},\
  }\bibfield  {title} {\enquote {\bibinfo {title} {Multiphase fluid dynamics
  simulations of product crossover in solar-driven, membrane-less water
  splitting},}\ }\href {\doibase 10.1016/j.xcrp.2021.100358} {\bibfield
  {journal} {\bibinfo  {journal} {Cell Rep. Phys. Sci.}\ }\textbf {\bibinfo
  {volume} {2}},\ \bibinfo {pages} {100358} (\bibinfo {year}
  {2021})}\BibitemShut {NoStop}%
\end{thebibliography}
\end{document}

\preprint{ }

\title{Supplementary Materials \\ \large Counter-balancing light absorption and ionic transport losses in the electrolyte for integrated solar water splitting with III-V/Si dual-junctions}
% Force line breaks with \\

\author{Moritz Kölbach}
 %\altaffiliation[Also at ]{Physics Department, XYZ University.}%Lines break automatically or can be forced %with \\
\affiliation{ 
Universität Ulm, Institute of Theoretical Chemistry,  Lise-Meitner-Str. 16, 89069 Ulm, Germany%\\This line break forced with \textbackslash\textbackslash
}
\author{Ciler Özen}
\affiliation{ 
Helmholtz-Zentrum Berlin für Materialien und Energie GmbH, Institute for Solar Fuels, Hahn-Meitner-Platz 1, 14109 Berlin, Germany %\\This line break forced with \textbackslash\textbackslash
}

\author{Oliver Höhn}%
\author{David Lackner}%
\author{Markus Feifel}%
\affiliation{ 
Fraunhofer Institute for Solar Energy Systems ISE, Heidenhofstraße 2, 79110 Freiburg, Germany %\\This line break forced with \textbackslash\textbackslash
}
\author{Fatwa F.~Abdi}%
\affiliation{ 
Helmholtz-Zentrum Berlin für Materialien und Energie GmbH, Institute for Solar Fuels, Hahn-Meitner-Platz 1, 14109 Berlin, Germany %\\This line break forced with \textbackslash\textbackslash
}
\author{Matthias M.~May}
 %\homepage{http://www.Second.institution.edu/~Charlie.Author.}
\affiliation{ 
Universität Ulm, Institute of Theoretical Chemistry,  Lise-Meitner-Str. 16, 89069 Ulm, Germany%\\This line break forced with \textbackslash\textbackslash
}

\renewcommand{\thepage}{S\arabic{page}}
\maketitle
%\linenumbers

\onecolumngrid

\renewcommand{\thetable}{S\arabic{table}}
\renewcommand{\thefigure}{S\arabic{figure}}
\renewcommand{\theequation}{S\arabic{equation}}

\textbf{Supplementary note 1: COMSOL Modelling}\\

The computational domain consisted of a stagnant liquid electrolyte channel of width $w_{cell}$ and height $d_{H_{2}O}$ between the two parallel anode electrodes of height $h_{anode}$ and a cathode of width $w_{cathode}$. In our study, $w_{cathode}$ was varied from 0.8 to 4\,cm, and $d_{H_{2}O}$ was varied from 0.1 to 1\,cm. The ratio of $h_{anode}$/$d_{H_{2}O}$ and $w_{cathode}$/$w_{cell}$ are kept constant at 0.8. Electrode thicknesses were neglected in our model (i.e. no ohmic substrate loss), and the electrode reactions were described on the electrode surfaces. The initial electrolyte concentration was 1\,M \ce{HClO4}. All other input parameters are listed in Table S2. COMSOL Multiphysics\textcopyright\,version 5.6 was employed to solve the steady-state governing transport (Nernst-Planck) and conservation equations of the ionic species, considering charge neutrality:
\begin{align}
-\nabla.N\textsubscript{i} + R\textsubscript{i} = 0 
\end{align}
\begin{align}
 N\textsubscript{i} = - D\textsubscript{i}\nabla c\textsubscript{i} - \frac{z\textsubscript{i} D\textsubscript{i}}{RT} F c\textsubscript{i} \nabla  \Phi\textsubscript{1}
\end{align}
Here, $N_{i}$ is the molar flux vector, $R_{i}$  is the reaction source term, $D_{i}$  is the diffusivity, $c_{i}$  is the concentration, and $z_{i}$ is the charge of species i (\ce{H+} or \ce{ClO4-}). $F$ is Faraday’s constant and  $\Phi_{l}$ is the electrolyte potential. The current density in the electrolyte ($j_{l}$) is calculated as follows:

\begin{align}
 j\textsubscript{l} = F \sum_{i} z\textsubscript{i} N\textsubscript{i} 
\end{align}

At the electrode surface, mass fluxes were determined by the local current density ($j_{loc}$) and the stoichiometry coefficients ($\nu_{i}$) of the redox equilibrium reaction.
\begin{align}
 R\textsubscript{i} = \frac{- \nu\textsubscript{i} j\textsubscript{loc}}{nF} 
\end{align}
The stoichiometry coefficient for \ce{H+} and the number of electrons ($n$) for reactions on the cathode (anode) are -2 (-4) and 2 (4), respectively. The local electrode current density ($j_{loc}$) was determined by anodic Tafel equation at the anode and by cathodic Tafel equation at the cathode.
\begin{align}
 j\textsubscript{loc} = j\textsubscript{0} \cdot 10^{\left( \frac{\eta}{b} \right)}
\end{align}
Here, $j_{0}$ is the exchange current density, $b$ is the Tafel slope and $\eta$ is the overpotential. The latter was determined for the anode and cathode by:
\begin{align}
 \eta\textsubscript{anode} = \Phi\textsubscript{s} - \Phi\textsubscript{l} - 1.23\,V + \frac{RT}{nF}  \ln{\left( \frac{c\textsubscript{H\textsuperscript{+}}}{c\textsubscript{H\textsuperscript{+}, bulk}} \right) }^{\nu_{i}}
\end{align}
\begin{align}
 \eta\textsubscript{cathode} = \Phi\textsubscript{s} - \Phi\textsubscript{l} - 0\,V + \frac{RT}{nF}  \ln{\left( \frac{c\textsubscript{H\textsuperscript{+}}}{c\textsubscript{H\textsuperscript{+}, bulk}} \right) }^{\nu_{i}} 
\end{align}
Here, $\Phi_{s}$ is the electrode surface potential.

Average current density was applied at the anode’s electrical contact. The potential of the cathode’s electrical contact was set to ground. On other boundaries insulation boundary condition ($-n.i_{l}$ = 0; $-n.i_{s}$ = 0) was considered. No flux boundary conditions ($-n._{J}$ = 0) were considered on all boundaries of cell for mass transport. A mesh independence study was carried out and 195695 triangular elements were used for a converging solution (for the cell sizes of $w_{cell}$ = 5\,cm and $d_{H_{2}O}$ = 1\,cm). Afterwards a parametric study was carried out for different cell sizes and current densities as illustrated in Figure S2 and in Figure 2c. For the solutions of the dependent variables, the PARDISO (parallel sparse direct solver) approach, which is a direct method based on lower-upper (LU) factorization (matrix triangulation), was used.\\
\\

\textbf{Supplementary note 2: Experimental solar cell input parameters}\\

The experimental IV and EQE characteristics of the AlGaAs//Si solar cell were implemented in the model as follows. A general open circuit voltage loss ($V_{oc, ~loss}$) is estimated from the experimental $V_{oc,~ref}$ of the AlGaAs//Si reference solar cell and the top ($E\textsubscript{g,~top, ~ref}$) and bottom ($E\textsubscript{g, bottom, ref}$) absorber bandgap of 1.75\,eV and 1.1\,eV, respectively: 
\begin{align}
V\textsubscript{oc, loss}  = (E\textsubscript{g, top, ref}  + E\textsubscript{g, bottom, ref}) - V\textsubscript{oc, ref} 
\end{align}

This voltage loss is assumed to be constant for dual-junction cells with different top absorber bandgaps and the $V_{oc}$ is estimated as follows:

\begin{align}
V\textsubscript{oc}   = (E\textsubscript{g, top}  + E\textsubscript{g, bottom}) - V\textsubscript{oc, loss} 
\end{align}

The temperature dependence of the open circuit voltage can be approximated as

\begin{align}
    V\textsubscript{oc}\left( T\textsubscript{device} \right) {=} V\textsubscript{oc, T\textsubscript{ref}} {+} T\textsubscript{coeff, V\textsubscript{oc}} {\cdot} \left( T\textsubscript{device} {-} T\textsubscript{ref} \right) {\cdot} V\textsubscript{oc, T\textsubscript{ref}}
\end{align}

Here, $T\textsubscript{coeff, V\textsubscript{oc}}$ is the relative open circuit potential temperature coefficient at the reference temperature $T\textsubscript{ref}$. The shape of the experimental IV-characteristics of the AlGaAs//Si cell is reproduced via empirical fitting according to the following formula\cite{Bellini_2009}:

\begin{align}
j = j\textsubscript{sh} {\cdot} \left[1 - C\textsubscript{1} {\cdot} \left( \exp\left( \frac{ V }{C\textsubscript{1} {\cdot} V\textsubscript{oc} } \right) - 1 \right)  \right]
\end{align}

Here, $j\textsubscript{sh}$ is the short current density, while $C\textsubscript{1}$ and $C\textsubscript{2}$ are empirical fitting parameters. To estimate the amount of parasitic absorption, recombination losses, and reflection, the total experimental EQE is multiplied with the solar spectra modified by the water layer. From these spectra, the $j\textsubscript{sh}$ for each top absorber bandgap is obtained. Together with the $V\textsubscript{oc}$, and the empirical fitting parameters $C\textsubscript{1}$ and $C\textsubscript{2}$, the IV characteristics for each top absorber bandgap is estimated using equation (4). Note that the temperature dependence of the short-circuit current density is neglected in the model. A detailed description of the full model can be found in the YaSoFo documentation.\cite{May_yasofo-data_2020}\\
\\

\begin{figure}[ht!]
\includegraphics[width=0.5\linewidth]{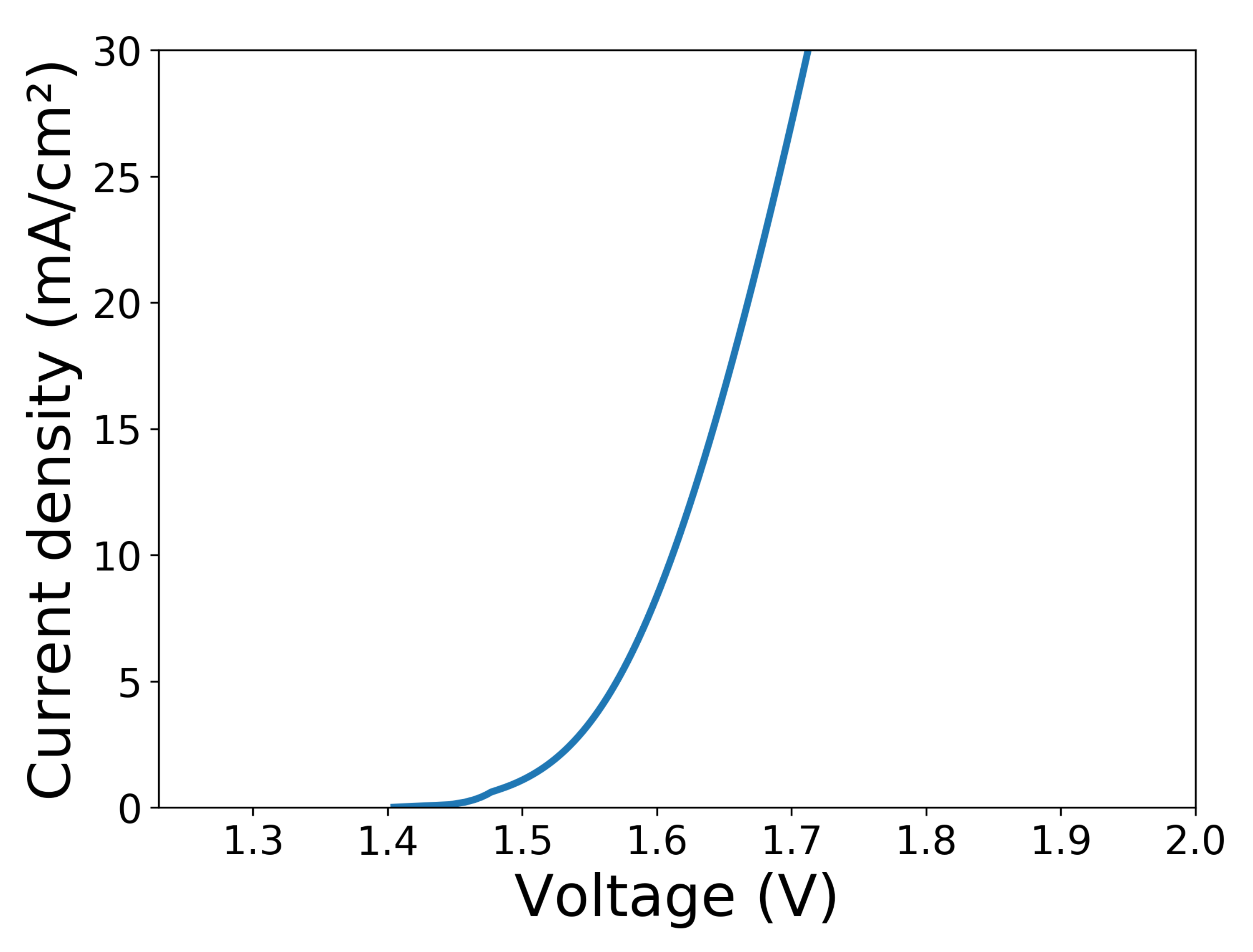}% Here is how to import EPS art
\caption{\label{fig:ohmic_transport} 2-electrode water splitting IV-characteristics resulting from the input parameters (see Table S1) used in Fig. 1 in the manuscript.}
\end{figure}

\begin{figure}[ht!]
\includegraphics[width=0.9\linewidth]{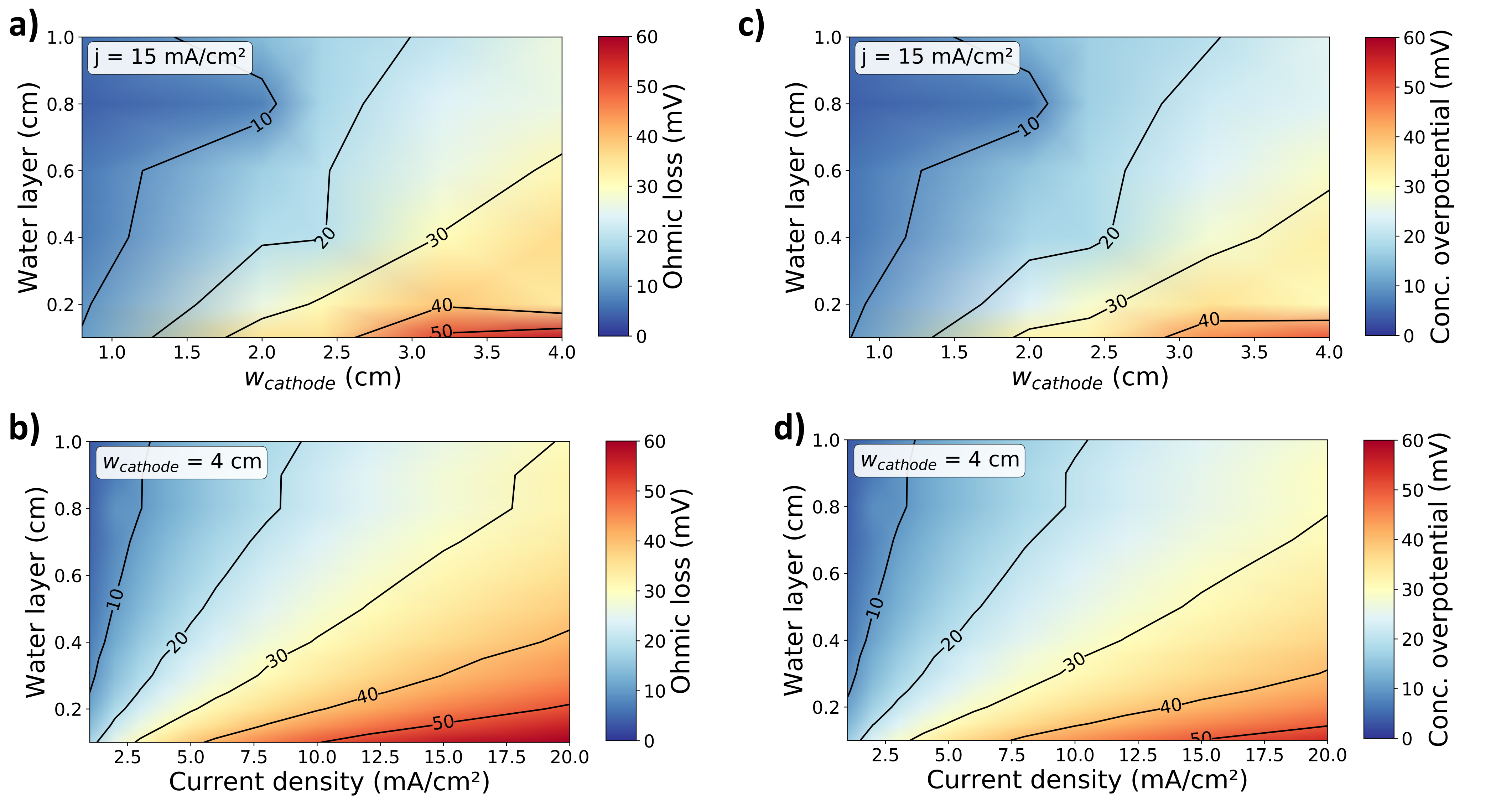}% Here is how to import EPS art
\caption{\label{fig:ohmic_transport}  Contributions of (a), (b) the ohmic and (c), (d) concentration overpotentials to the total voltage losses shown in Figs. 2(b) and 2(c) in the manuscript simulated with COMSOL.}
\end{figure}

\begin{figure}[ht!]
\includegraphics[width=0.5\linewidth]{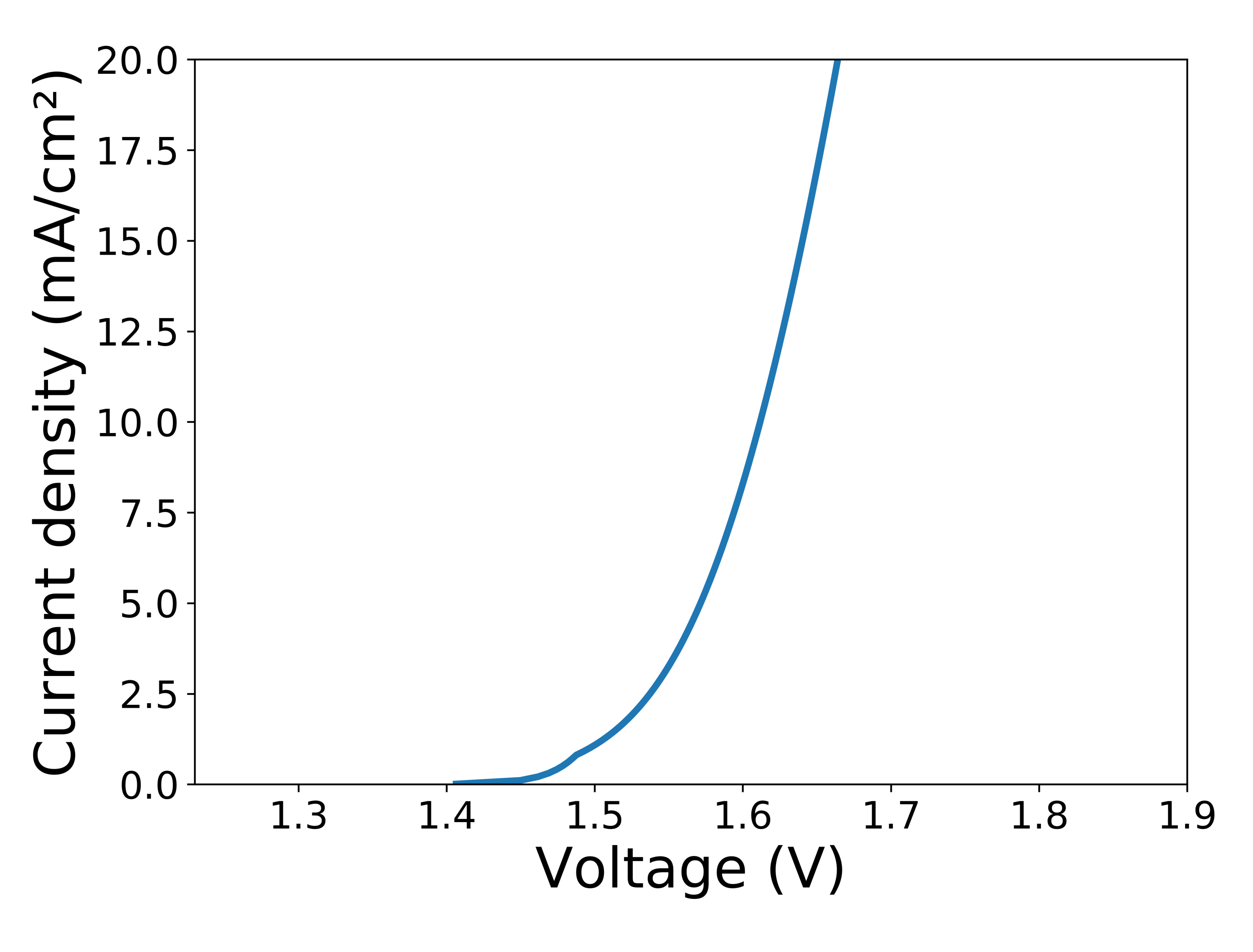}% Here is how to import EPS art
\caption{\label{fig:ohmic_transport} 2-electrode water splitting IV-characteristics resulting from the input parameters (see Table S1) used in Figs. 3(c) and 3(e) in the manuscript. }
\end{figure}

\begin{figure}[ht!]
\includegraphics[width=0.5\linewidth]{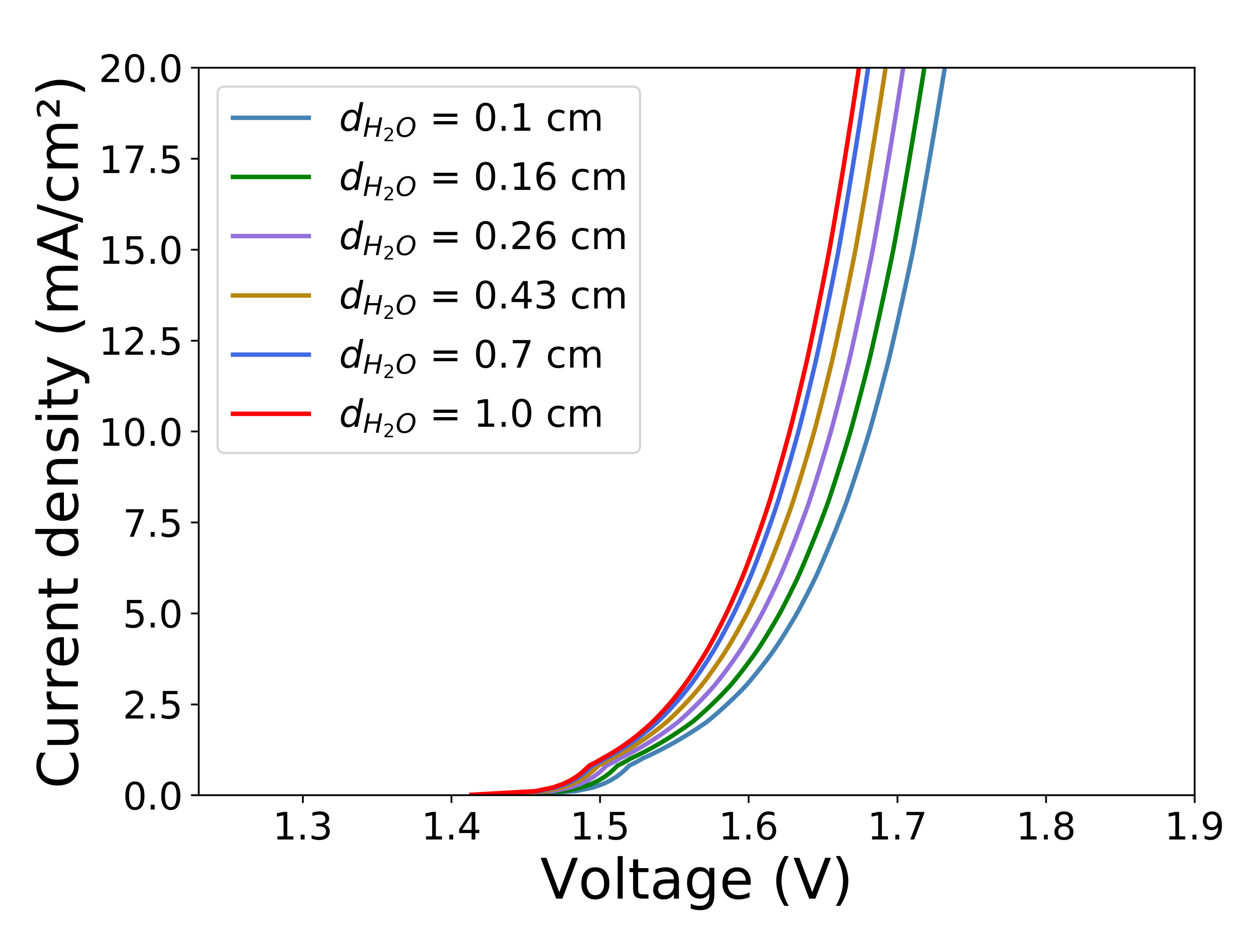}% Here is how to import EPS art
\caption{\label{fig:ohmic_transport} 2-electrode water splitting IV-characteristics for several water layer thicknesses resulting from the input parameters (see Table S1) and the voltage losses (see Fig. 2(c)) used in Figs. 3(e) and 3(f) in the manuscript.}
\end{figure}

\clearpage

\begin{table}[h!]
\begin{center}
    \caption{YaSoFo model input parameters.}
    \label{table:2}
    \begin{tabular}{|p{2.9cm}|p{6.5cm}|p{3.1cm}|p{2cm}|}
    \hline
    $j\textsubscript{0, OER}$\newline  (Fig. 1) & OER exchange current density  & $1.86\times{10^{-9}}$\,A/cm$^2$ for IrO$_x$ at $T\textsubscript{ref}$\,=\,323\,K & \cite{Lettenmeier_2016} \\ \hline
    $j\textsubscript{0, OER}$\newline (Fig. 3) & OER exchange current density  & $5\times{10^{-10}}$\,A/cm$^2$ at $T\textsubscript{ref}$\,=\,323\,K & - \\ \hline
    $E\textsubscript{a, OER}$ & Activation Energy OER & 52\,kJ/mol & \cite{Lettenmeier_2016} \\ \hline
    $\alpha\textsubscript{a,OER} \cdot z$ &  Anodic charge transfer coefficient of the OER multiplied with the electrons involved (can be extracted from the Tafel slope) & 1.5 & \cite{Lettenmeier_2016} \\ \hline 
    $j\textsubscript{0, HER}$ & HER exchange current density & $0.68\times{10^{-3}}$\,A/cm$^2$ for Pt at $T\textsubscript{ref}$~=~303\,K & \cite{Markovic_1997} \\ \hline
    $E\textsubscript{a, HER}$ & Activation Energy HER  & 13.2\,kJ/mol & \cite{Markovic_1997} \\ \hline
    $\alpha\textsubscript{c, HER} \cdot z$ &  Cathodic charge transfer coefficient of the HER multiplied with the electrons involved (can be extracted from the Tafel slope) & 1.2 & \cite{Markovic_1997} \\ \hline
    $T\textsubscript{device}$\newline (Fig. 1) & Device Temperature & 25\,$^\circ$C & - \\ \hline
    $T\textsubscript{device}$\newline (Figs. 3 and 4) & Device Temperature & 40\,$^\circ$C & - \\ \hline
    $R\textsubscript{cell}$\newline (Fig. 1) &  Ohmic cell resistance (Corresponds to two parallel electrodes with a distance of 1\,cm  and 1 M\,\ce{HClO4} as an electrolyte. The conductivity of 1\,M \ce{HClO4} is obtained via $\kappa\ {=} y\textsubscript{0} {+} m {\cdot} T\textsubscript{device} $ with y\textsubscript{0}~=~-1.00802\,S/cm and m~=~0.00453\,S/(cm${\cdot}$K); see reference\,\cite{Brickwedde_HClO4}) & 2.92\,$\Omega {\cdot} cm^{2} $\newline (T\textsubscript{device} = 25\,$^\circ$C) & - \\ \hline 
    $R\textsubscript{cell}$\newline (Fig. 3c and 3d) &  Ohmic cell resistance & 2.44\,$\Omega {\cdot} cm^{2} $\newline (T\textsubscript{device} = 40\,$^\circ$C) & - \\ \hline
    $T\textsubscript{coeff, V\textsubscript{oc}}$ &  Open circuit potential temperature coefficient of a double junction solar cell at $T\textsubscript{ref}$~=~25\,$^\circ$C & -0.3\,\%/K & - \\ \hline
    $n\textsubscript{i}$\newline (Fig. 1) & Diode ideality factor  & 1 & - \\ \hline
    Spectrum & Solar spectrum & AM1.5G\newline (ASTM G-173) & \cite{Gueymard_2004}\\ 
    \hline
    \end{tabular}
\end{center}
\end{table}

\clearpage

\begin{table}[h!]
\begin{center}
    \caption{Baseline parameters used in the voltage loss simulations.}
    \label{table:2}
    \begin{tabular}{|p{2.9cm}|p{6.5cm}|p{3.1cm}|p{2cm}|}
    \hline
    $c\textsubscript{\ce{H+}, i}$ & Initial \ce{H+} concentration & 1\, M & - \\ \hline
    $c\textsubscript{\ce{ClO4-}, i}$ & Initial \ce{ClO4-} concentration & 1\, M & - \\ \hline
    $D\textsubscript{\ce{H+}}$ & Diffusivity of \ce{H+}  & $9.3 \times 10^{-9} m^{2}$/s & \cite{Haynes_2017, US_Geo} \\ \hline
    $D\textsubscript{\ce{ClO4-}}$ & Diffusivity of \ce{ClO4-}  & $1.8 \times 10^{-9} m^{2}$/s & \cite{Buffle_2007} \\ \hline
    $T$ & Temperature & 313\,K & - \\ \hline
    $j\textsubscript{0, OER}$ & OER exchange current density at 313\,K & $1.0\times{10^{-9}}$\,A/cm$^2$ & \cite{Lettenmeier_2016} \\ \hline
    $j\textsubscript{0, HER}$ & HER exchange current density at 313\,K & $8.05\times{10^{-4}}$\,A/cm$^2$ & \cite{Markovic_1997} \\ \hline
    $b\textsubscript{OER}$ & OER Tafel slope at 313\,K & 41.3\,mV/dec & \cite{Lettenmeier_2016} \\ \hline
    $b\textsubscript{HER}$ & HER Tafel slope at 313\,K & 51.7\,mV/dec & \cite{Markovic_1997} \\ 
    \hline
    \end{tabular}
\end{center}
\end{table}

Note that because of the smallest considered water layer thickness of 0.1 cm in our calculations, effects of the Helmholtz-layer and the diffuse layer on transport and the optical properties of the device stack are neglected: They have thicknesses in the \AA ngstrom and nanometer range, respectively, and are not expected to significantly change the photocurrent. Furthermore, changes in reflectivity, that might be induced at the solid-liquid phase boundary by the strong electric fields in the Helmholtz-layer are not considered.

%\textbf{References}\\

%\bibliography{004_Si}

\providecommand{\noopsort}[1]{}\providecommand{\singleletter}[1]{#1}%